\pdfoutput=1
\documentclass[aps,prb,reprint,amsmath,amssymb,superscriptaddress]{revtex4-1}
\usepackage{bm}
\usepackage{graphicx}
\usepackage[colorlinks]{hyperref}
\usepackage{mathrsfs}
\usepackage{times}
\usepackage[dvipsnames]{xcolor}
\usepackage{float}
\usepackage{color}
\usepackage{mathtools}

\newcommand{\bit}{\begin{enumerate}}
	\newcommand{\eit}{\end{enumerate}}

\definecolor{bananayellow}{rgb}{1.0, 0.88, 0.21}
\definecolor{straw}{rgb}{0.32, 0.28, 0.1}

\begin{document}
	\title{Anomalous Random Multipolar Driven Insulators}
	\author{Hongzheng Zhao}
	\affiliation{Max-Planck-Institut f\"ur Physik komplexer Systeme, N\"othnitzer Stra\ss e 38, 01187 Dresden, Germany}
	\affiliation{Blackett Laboratory, Imperial College London, London SW7 2AZ, United Kingdom}
	\author{Mark S. Rudner}
	\affiliation{Center for Quantum Devices and Niels Bohr International Academy,
		University of Copenhagen, 2100 Copenhagen, Denmark}
	\affiliation{Department of Physics,
		University of Washington, Seattle, Washington, USA}
	\author{Roderich Moessner}
	\affiliation{Max-Planck-Institut f\"ur Physik komplexer Systeme, N\"othnitzer Stra\ss e 38, 01187 Dresden, Germany}
	\author{Johannes Knolle  }
	\affiliation{Department of Physics TQM, Technische Universit{\"a}t M{\"u}nchen, James-Franck-Stra{\ss}e 1, D-85748 Garching, Germany}
	\affiliation{Munich Center for Quantum Science and Technology (MCQST), 80799 Munich, Germany}
	\affiliation{Blackett Laboratory, Imperial College London, London SW7 2AZ, United Kingdom}
	\begin{abstract}
		
		It is by now well established that periodically driven quantum many-body systems can realize topological non-equilibrium phases without any equilibrium counterpart. Here we show that even in the absence of time translation symmetry, non-equilibrium topological phases of matter can exist in aperiodically driven systems for tunably parametrically long prethermal lifetimes. As a prerequisite, we first demonstrate the existence of long-lived prethermal Anderson localization in two dimensions under random multipolar driving. We then show that the localization may be topologically non-trivial with a quantized bulk orbital magnetization even though there are no well-defined Floquet operators. We further confirm the existence of this {\it Anomalous Random Multipolar Driven Insulator} by detecting quantized charge pumping at the boundaries, which renders it experimentally observable.

	\end{abstract}
	\maketitle
	
	\section{Introduction}
	The exploration of non-equilibrium many-body phenomena has flourished in the past decades as time-dependent driving opens new pathways for controlling  quantum systems~\cite{arute2019quantum,gross2017quantum,martinez2016real,maczewsky2017observation,yang2020observation}. Beyond the research programme of ``Floquet-engineering'' sought-after equilibrium phases of matter~\cite{schweizer2019floquet,oka2019floquet}, of particular interest is the possibility to realise intrinsically dynamical phases without direct analogies in static systems~\cite{kitagawa2010topological,khemani2016phase}. For example, combining the discrete time translational symmetry (TTS) of periodically driven systems with many-body localization (MBL)~\cite{parameswaran2018many} may stabilize a discrete time crystal phase spontaneously breaking  TTS~\cite{moessner2017equilibration,else2020discrete,khemani2019brief,pizzi2021higher}.  
	Floquet systems may also host a variety of topological non-equilibrium  phases such as the two-dimensional (2D) anomalous Floquet Anderson insulators (AFAI)~\cite{titum2016anomalous,rudner2013anomalous} and its interacting generalization, the Anomalous Floquet insulator (AFI)~\cite{nathan2019anomalous}. Although  fully localized in the bulk, they support quantized chiral edge currents robust to generic perturbations.
	
	It is natural to ask whether dynamical phases may exist beyond the Floquet paradigm~\cite{dumitrescu2018logarithmically,friedman2020topological,zhao2019floquet,else2020long,long2021many,martin2017topological,nathan2020quasiperiodic}. 
	Recently the fate of Floquet topological edge states in the presence of white noise has been studied~\cite{rieder2018localization,timms2020quantized}.
	Here we will address the question: can topological phases exist in aperiodically driven systems for tunably parametrically long times when the temporal aperiodicity is strong?
	If so, is there a diagnostic capable of capturing the topological nature of the system once TTS is explicitly broken? At first sight it seems unlikely because the absence of TTS precludes the usual definition of Floquet operators. Therefore, the topological characterization, e.g., via the bulk winding numbers of the AFAI Floquet states, cannot be applied~\cite{titum2016anomalous,crowley2019topological}. In addition, the stability of MBL in Floquet systems is essential for the realization of the AFAI/AFI. In contrast, MBL is unstable for aperiodic drives and, thus, cannot prevent heating to a featureless infinite temperature state~\cite{dumitrescu2018logarithmically,long2021many,zhao2021localization}. 
	
	Nevertheless, we here provide an affirmative answer to the above question by constructing a concrete example.  To do so, 
	we introduce an aperiodic step-wise driving scheme which extends the Floquet protocol for the realization of the AFAI~\cite{titum2016anomalous,rudner2013anomalous} to $n-$random multipolar driving ($n-$RMD) as recently proposed in Ref.~\onlinecite{zhao2021random}. A key observation for our purpose is that for generic (non-integrable) many-body systems a transient but long-lived prethermal steady state emerges in $n-$RMD drives, whose lifetime $\tau$ scales universally as $\tau \propto (1/T)^{2n+1}$ for finite $n$~\cite{zhao2021random,mori2021rigorous} (with $T$ the duration of the fundamental time evolution block as introduced below). The lifetime grows faster than any power law in $1/T$ in the $n\to \infty$ limit, where $n-$RMD corresponds to the quasiperiodic Thue-Morse (TM) sequence~\cite{nandy2017aperiodically}.  Here, we first establish that disorder induced localization also follows this scaling and can indeed persist as a long-lived prethermal phenomenon~\cite{kuwahara2016floquet,abanin2017rigorous,machado2020long,pizzi2021classical} before the system eventually  delocalizes. Next we show how long-lived  localization can lead to a prethermal topologically nontrivial 2D insulator, which we dub the {\it Anomalous Random Multipolar Driven Insulator} (ARMDI). 
	
	Crucially, although TTS is absent, we can show that the bulk orbital magnetization density~\cite{nathan2017quantized} remains quantized over the prethermal time scale. We furthermore confirm the existence of the ARMDI by detecting the quantized charge pumping at its boundaries. Although we concentrate on the non-interacting limit, the prethermal lifetime of the ARMDI strikingly scales in the same way as established for  non-integrable systems in 1D~\cite{mori2021rigorous,zhao2021random}, suggesting that the ARMDI remains robust with respect to the addition of sufficiently weak interactions.
	
	\section{The driving protocol}
	\begin{figure}[t]
		\centering
		\includegraphics[width=\linewidth]{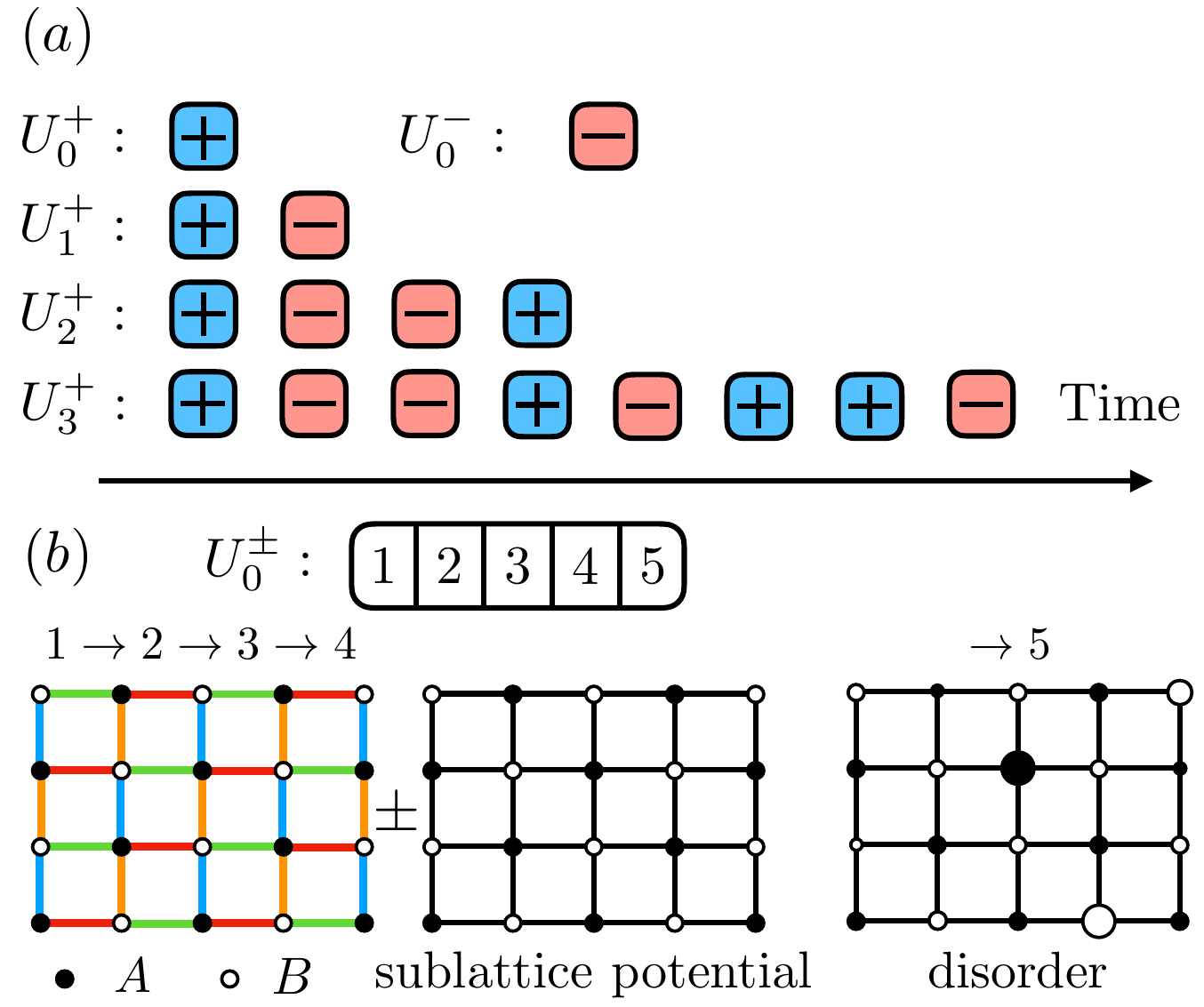}
		\caption{Scheme of the driving protocol. (a) Sequence of driving blocks $U_0^{\pm}$ employed to realize the TM protocol up to time $2^nT$. (b) The elementary time evolution operator $U_0^{\pm}$ involves five driving steps. The first four steps contain hopping processes on different bonds of different colors, meanwhile a sublattice potential of amplitude $\pm\delta h$ is applied. The disorder potential appears in the fifth step, leading to Anderson localization.
		}
		\label{fig:protocol}
	\end{figure}
	
	We first introduce the protocol, starting with a system of non-interacting spinless fermions on a 2D bipartite lattice driven by the step-wise time-dependent Hamiltonian
	$
	H(t)= H_{\mathrm{id}}(t)+H_{\mathrm{dis}}(t),$
	which is periodic in time: $H(t)=H(t+T)$ with period $T$. Within a period, the first term consists of translationally invariant hopping terms described by~\cite{titum2016anomalous}
	\begin{eqnarray}
		\label{eq.identity}
		H_{\mathrm{id}}(t)=J\sum_{\mathbf{r} \in A} \sum_{n=1}^{4} f_{n}(t)\left(c_{\mathbf{r}+\mathbf{d}_{n}}^{\dagger} c_{\mathbf{r}}+\mathrm{h.c.}\right),
	\end{eqnarray}
	where $f_n(t)=1$ for $(n-1)T/5\leq t<nT/5$ and zero otherwise. The summation over $\mathbf{r}$ is performed over all sites of sublattice A as illustrated in Fig.~\ref{fig:protocol} (b). The vectors $\{\mathbf{d}_n\}$ are defined as $\mathbf{d}_1=-\mathbf{d}_3=(0,1)$ and $\mathbf{d}_2=-\mathbf{d}_4=(1,0)$. This model has a solvable limit when the hopping amplitude $J$ is chosen as $JT/5 = \pi/2$ 
	such that one particle prepared on site $\mathbf{r}$ can be completely transferred to site $\mathbf{r}+\mathbf{d}_n$ over step $n$ of the cycle. The fifth step of the cycle involves the disorder potential
	$
	H_{\mathrm{dis}}(t)= f_{5}(t)\sum_{\mathbf{r}} h_{\mathbf{r}} c_{\mathbf{r}}^{\dagger} c_{\mathbf{r}}$,
	where $h_{\mathbf{r}}$ is uniformly chosen in the interval $[-h_{\mathrm{max}}, h_{\mathrm{max}}]$. This Floquet model hosts the topologically protected non-equilibrium phase, AFAI, which features  chiral edge modes together with a fully localized bulk~\cite{titum2016anomalous}. It remains stable to generic time-periodic perturbations (with the same period as the drive, $T$) as long as no topological phase transition occurs.
	
	Building on this foundation, we now study an {\it aperiodically} driven cousin of the AFAI -- the ARMDI introduced above -- and show that nontrivial topology persists as a new long-lived prethermal phase.
The aperiodic driving protocol is defined in terms of sequences which toggle in an irregular fashion between two types of evolution blocks, labeled by $+$ and $-$. The corresponding stroboscopic time evolution operators for these blocks are defined as
\begin{eqnarray}
	\label{eq.operatorphysical}
	U_0^{\pm}=\mathcal{T} e^{-i \int_{0}^{T} d s {H}^{\pm} (t)},
\end{eqnarray} where the Hamiltonian reads $H^\pm(t) = H_{\mathrm{id}}(t)+H_{\mathrm{dis}}(t)+H_{\mathrm{pert}}^{\pm}(t)$ with a local perturbation $H_{\mathrm{pert}}^{\pm} (t)$. 
Now $T$ defines the duration of the fundamental evolution block, and its inverse denotes the characteristic driving rate of the protocol.
The qualitative behavior of the ARMDI does not depend on the specific form of the perturbation but for concreteness we consider the perturbation of the form
$
H_{\mathrm{pert}}^{\pm} (t) = (\delta J/J)H_{\mathrm{id}}(t){\pm} \left[1-f_5(t)\right]\delta hH_{\mathrm{sub}}.
$
The first term modifies the hopping amplitudes in the ``ideal'' hopping cycle by an amount $\delta J$, which breaks the perfect transfer of particles between sites on each step. 
The second term is a time-dependent sublattice potential, $H_{\mathrm{sub}} =  (\sum_{\mathbf{r}\in A}n_{\mathbf{r}}-\sum_{\mathbf{r}\in B}n_{\mathbf{r}})$, with opposite signs in the two types of evolution blocks, see Fig.~\ref{fig:protocol} (b), which is nonzero during all but the fifth step.  For any $\delta h \neq 0$, the stroboscopic block evolution operators defined in Eq.~\eqref{eq.operatorphysical}  do not commute: $[U_0^{+},U_0^-]\neq0$. Thus the driving with $\delta h \neq 0$ is qualitatively different from that of the
aforementioned Floquet drive. 

We consider two types of aperiodic driving protocols where TTS is explicitly broken. 
The first protocol is the quasiperiodic TM sequence shown in Fig.~\ref{fig:protocol} (a). In this case, the time evolution operator $U_{n}^+$ at time $t=2^n T$ is constructed recursively from the elementary block evolution unitaries $U_0^{\pm}$ [Eq.~(\ref{eq.operatorphysical})] as 
$
U_{n}^{\pm}=U_{n-1}^{\mp} U_{n-1}^{\pm}$~\footnote{\label{TM} An important feature of the TM protocol is that, due to its recursive nature, only a linearly increasing number of matrix multiplications is required to obtain the result of an exponentially long time evolution.}.
The other protocol we consider is the $n-$RMD protocol.
For $n$-RMD, at each iteration one of the two $n-$th order multipolar operators $U_n^{\pm}$ defined above is randomly chosen to propagate the state~\cite{zhao2021random}.
In both cases, it is clear that Floquet theory manifestly does not apply.
Importantly, this means that the eigenstates and quasi-energies that underpin the topological classification of
Floquet systems are absent and a different approach is needed. 

\section{Prethermal localization}
For generic non-integrable systems subjected to $n-$RMD or TM driving, 
it was both rigorously shown and numerically verified that long-lived prethermal states form in the rapid driving regime, i.e., when the inverse of the corresponding fundamental block duration, $1/T_{\mathrm{RMD}}$ or $1/T_{\mathrm{TM}}$, respectively, is the dominant energy scale of the system~\footnote{The fundamental block duration for $n$-RMD or TM driving is defined as the common duration of the evolution blocks $U^\pm_0$, which we assume are equal in length. The lifetime $\tau$ of the 
	prethermal state scales algebraically for $n-$RMD as  $\tau_{\mathrm{RMD}}\sim T_{\mathrm{RMD}}^{-(2n+1)}$, while for TM driving the lifetime scales as $\tau_{\mathrm{TM}} \sim e^{C[\ln (T_{\mathrm{TM}}^{-1}/g)]^{2}}$ with a constant $C$ and a typical local energy scale $g$~~\cite{zhao2021random,mori2021rigorous}.}.
For the model considered here, it is not {\it a priori} clear whether the same phenomenology holds, because Anderson localization might drastically change the system's approach to eventual equilibration. 
Also, to remain close to the ``ideal hopping'' condition, i.e., $JT/5 = \pi/2$, the hopping amplitude $J$ must increase proportionally with the driving rate $1/T$.
Hence, the requirement of rapid driving cannot be satisfied.

To demonstrate that the strongly-driven, disordered model that we consider also exhibits long-lived prethermalization, we employ a unitary transformation $\label{eq:Q}Q(t)=\mathcal{T} e^{-i \int_{0}^{t} d s H_{\mathrm{id}}(s)}$  that removes the evolution due to the ``ideal'' part of the hopping~\cite{nathan2019anomalous}.
The Hamiltonian in the corresponding rotating frame reads $\tilde{H}(t)=Q^{\dagger}(t) \hat{H}(t) Q(t)-i Q^{\dagger}(t) \dot{Q}(t)$. Note that this transformation is periodic in time: $Q(t)=Q(t+T)$. 
In the rotating frame, the energy scale $J$ does not contribute to the norm of the Hamiltonian, see Supplementary Material (SM).
A rapid driving regime is hence achieved for $h_{\mathrm{max}},\delta J,\delta h\ll T^{-1}$, where a long-lived prethermal localization can be established. 
The localization length in the rotating frame increases with the ratio $\delta J/h_{\mathrm{max}}$; the value of 
$\delta h$ determines the strength of the temporally aperiodic portion of the drive, which eventually delocalizes the system. In fact, due to the periodicity of $Q(t)$, the stroboscopic time evolution operators in the physical and rotating frames coincide. Therefore, rigorous results obtained in the rotating frame for $n-$RMD and TM driving directly remain valid in the physical frame at stroboscopic times.

\begin{figure}
	\centering
	\includegraphics[width=\linewidth]{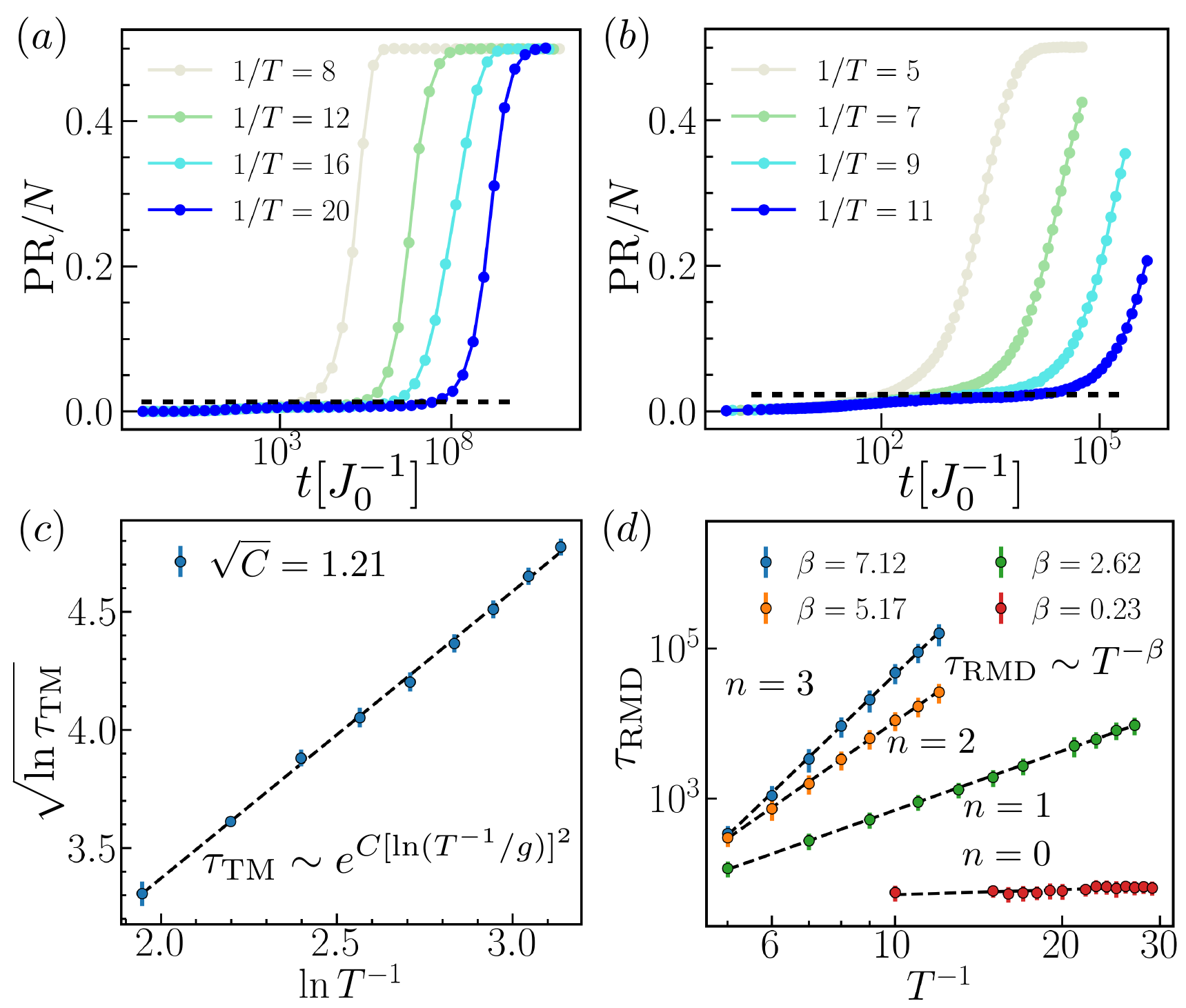}
	\caption{The participation ratio for (a) TM driving and (b) 3-RMD, with varying driving rates, quickly saturates to a prethermal plateau before increasing to 0.5 (indicating delocalization at long times). The dependence of the prethermal lifetime on the driving rate is shown for (c) TM driving and (d) the $n$-RMD drivings. We use parameters $\delta J=1.2J_0, h_{\mathrm{max}}=6J_0,\delta h=7J_0$ and system size $70\times 70$ and $40\times 40$ for TM driving and RMD respectively. $T^{-1}$ is in units of $J_0$.}
	\label{fig:iprdynamics}
\end{figure}

In the following, we numerically verify the existence of prethermal Anderson localization as a prerequisite for the ARMDI. We consider a lattice of $N = L_x \times L_y$ sites with periodic boundary conditions, subjected to the TM and $n-$RMD protocols described above.
To demonstrate the existence of prethermal localization, we  prepare an initial state of a single particle on site $m$ and quantify the degree of localization over time using the participation ratio (PR) 
$
\mathrm{PR}_{m}(t)=1 / \sum_{j=1}^{N}\left|\psi_{j}^{m}\right|^{4},
$
where
${\psi^m_j(t)}$ defines the single particle wavefunction on site $j$ at time $t=MT$, and $M$ denotes the total number of fundamental evolution blocks $U_0^{\pm}$. 
We also average over all possible initial states to obtain ${\mathrm{PR}} = \langle \mathrm{PR}_{m}\rangle$. For localized states, the participation ratio scales as $\mathrm{PR}/N\sim1/N$, whereas for delocalized states one has $\mathrm{PR}/N\sim\mathcal{O}(1)$~\cite{janssen1998statistics}.

The evolution of $\mathrm{PR}/N$ for the TM and $3-$RMD protocols are plotted in Fig.~\ref{fig:iprdynamics} (a) and (b), respectively. 
As the particle's wave function locally spreads,  $\mathrm{PR}/N$ first increases and saturates to a small value (black dashed line) at time $t\sim10^2J_0^{-1}$. 
(Here $J_0$ is a reference energy scale that we use for the scaling of all numerical parameters in this work.)
The initial rise of $\mathrm{PR}/N$ is independent of the driving rate $1/T$ as long as it is large.
$\mathrm{PR}/N$ remains nearly constant in the prethermal regime, confirming the existence of long-lived prethermal localization. Only after a large time window, 
$\mathrm{PR}/N$ rapidly increases to the eventual plateau at the value 0.5, corresponding to a final steady state evenly occupying the entire space. 

For both types of driving, the prethermal lifetime of localization increases with $1/T$. 
To enable the numerical extraction of a prethermal lifetime, we first define a time $t_x$ such that $\mathrm{PR}(t_x)/N=x$. 
Since the choice of $x$ is somewhat arbitrary, 
we define the prethermal lifetime as the average $\tau=\langle t_x\rangle_x$ performed over five threshold values $x=x_0,x_0\pm\epsilon,x_0\pm\epsilon/2$. For TM driving,  $x_0=0.2,\epsilon=0.06$ and for RMD, we use $x_0=0.05,\epsilon=0.03$~\footnote{The scaling behavior does not show qualitative dependence on the
	precise values  $x$ of the threshold, as long as the state is in the prethermal regime at the corresponding time $t_x$. }.
The dependence of $\tau$ on $1/T$ is depicted in Fig.~\ref{fig:iprdynamics} (c) and (d) respectively for TM and RMD protocols.
As shown in Fig.~\ref{fig:iprdynamics} (c), the numerical results fit well with the analytical prediction of $\tau_{\mathrm{TM}}\sim e^{C[\ln (T^{-1}/g)]^{2}}$ for TM driving~\cite{mori2021rigorous}, where the slope defines the constant $\sqrt{C}$. 
In contrast, a log-log scale is used in Fig.~\ref{fig:iprdynamics} (d) where the linear fit indicates that $\tau_{\mathrm{RMD}}$ scales algebraically with the driving rate as $\tau_{\mathrm{RMD}}\sim T^{-\beta}$ for $n-$RMD. The fitted exponent reads approximately $\beta\approx 2n+1$ for $n\geq 1$, again in accordance with rigorous predictions~\cite{zhao2021random,mori2021rigorous}. For the purely random driving with $n=0$, the system always quickly delocalizes around $t\sim10J_0^{-1}$ and prethermal localization cannot be established. 

\begin{figure}
	\centering
	\includegraphics[width=\linewidth]{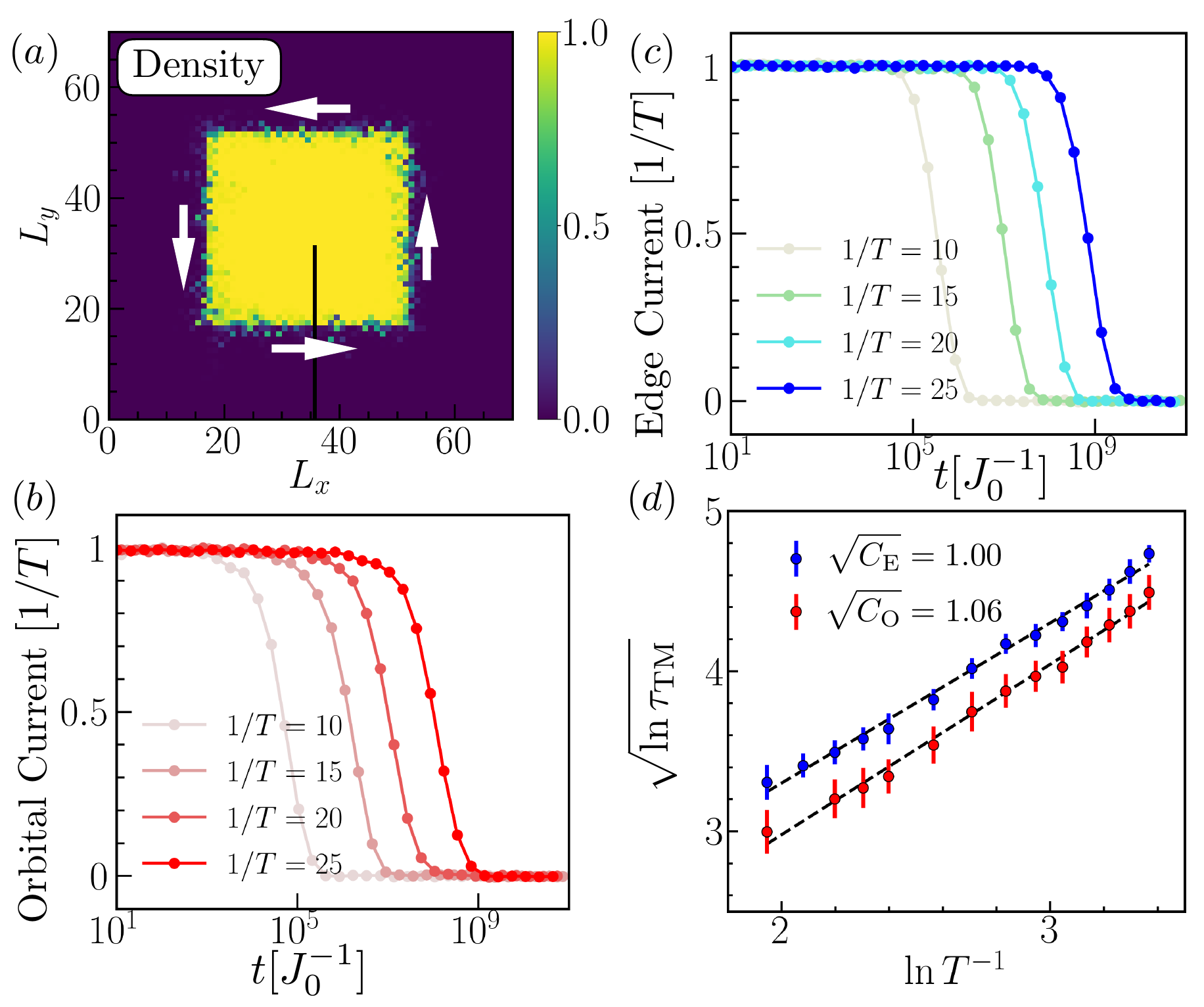}
	\caption{ Dynamics and scaling of prethermal lifetime for TM driving. (a) Particle density for a single random realization in the prethermal regime at time $t\approx10^3J_0^{-1}$ for $1/T=20J_0$. (b) Quantized orbital current serving as a prethermal topological order parameter. (c) Chiral edge current at the boundary between filled and empty sites. (d) Scaling of the lifetime of the quantized current. We use parameters $\delta J=1J_0,h_{\mathrm{max}}=20J_0,\delta h=7J_0.$ $T^{-1}$ is in units of $J_0$.}
	\label{fig:current}
\end{figure}

\section{Quantized charge pumping}We now confirm that the prethermal localized phase
is topological and that its topological nature is characterized by a prethermal bulk invariant. As the system possesses neither a Floquet spectrum nor a periodic micromotion operator, here we diagnose nontrivial topology using the time-averaged orbital magnetization density as introduced in Ref.~\onlinecite{nathan2017quantized} and detailed in the SM.

For the AFAI, which is stable in the long-time limit, the magnetization density averaged over an infinitely long time window, $\bar{m}_{\infty}$, was shown to be a topological invariant in units of $1/T$~\cite{nathan2017quantized}. This quantity is obtainable by using a finite ``droplet" 
constructed such that its interior is completely filled with particles whereas its surroundings are unoccupied. 
According to Ampere's law, for a droplet of sufficiently large size compared with the localization length, and in a stationary state, 
the magnetization density deep inside the droplet 
equals the time-averaged orbital current flowing at the droplet's boundary~\cite{nathan2017quantized}.

As localization in our system has a finite lifetime, instead of employing $\bar{m}_{\infty}$, we here use the magnetization density $\bar{m}_{T}(t)$ averaged over each block of duration $T$. Although $\bar{m}_{T}(t)$ is time-dependent, we will show that it remains approximately constant and quantized in units of $1/T$ in the prethermal regime. 

To extract $\bar{m}_{T}(t)$, we initially fill a square droplet ($35\times35$) centered in the middle of the square lattice ($70\times 70$) with periodic boundary conditions. Subject to TM driving, the droplet starts to evolve and remains well localized during the prethermal regime. As shown in Fig.~\ref{fig:current} (a), for a single disorder realization, the particle density at time $t\approx10^3J_0^{-1}$ 
remains close to the initial distribution with a slightly broadened boundary.
The orbital current $\bar{I}_C(t)$ can be obtained by integrating the expectation value of the current operator $I_C = -i\sum_{\mathbf{r} \in \mathrm{D}}\left[J_{\mathbf{r} \mathbf{r}'}(t) {c}_{\mathbf{r}}^{\dagger} {c}_{\mathbf{r}'}-J_{\mathbf{r}' \mathbf{r}}(t) {c}_{\mathbf{r}'}^{\dagger} {c}_{\mathbf{r}}\right]$ over a complete evolution block of duration $T$, 
where the set $\mathrm{D}$ includes all sites along one side of the cut [solid black line in Fig.~\ref{fig:current} (a)] and $J_{\mathbf{r} \mathbf{r}'}(t)$ is the time-dependent hopping to the adjacent sites on the other side of the cut, as defined in Eq.~\ref{eq.identity}. The average over different cuts is performed to reduce spatial and temporal fluctuations.
Clearly, as shown in Fig.~\ref{fig:current} (b), a prethermal plateau in the orbital current can be identified at the integer value $\bar{I}_C=1/T$. The current drops to zero only after a parametrically long time scale which substantially increases for larger $1/T$. Using the Ampere's law as discussed above, on the prethermal plateau we extract $\bar{m}_T(t)=\bar{I}_T(t)$.
Consequently, this defines the prethermal topological bulk invariant $\nu(t)\coloneqq T\bar{m}_T(t)$ and suggests the existence of a prethermal topologically non-trivial ARMDI. 

This is further verified by confirming the existence of a robust chiral edge current circulating at the boundary between filled and empty sites, in coexistence with a fully localized bulk.
We now consider the lattice geometry of a cylinder of size $70\times 70$ where the upper half is occupied. 
The edge current averaged over each evolution block of length $T$ starting at $t=2^n T$ is plotted in Fig.~\ref{fig:current} (c). 
The pumped charge per evolution block remains at the quantized value in the prethermal regime, before decaying when delocalization sets in.
Similar to the lifetime of prethermal localization, we define the lifetime $\tau$ of both the prethermal orbital and edge current by using the threshold values 0.6/T for the current~\footnote{Numerical noise can be suppressed in terms of an average over the threshold values $0.6/T,0.6\pm0.3/T,0.6\pm0.15/T$}. As presented in Fig.~\ref{fig:current} (d) {for TM driving}, their relation with the driving rate again fits well with the scaling $\tau_{\mathrm{TM}}\sim e^{C[\ln (T^{-1}/g)]^{2}}$. Both the fitted slopes give approximately the same exponent $\sqrt{C_{\mathrm{E}}},\sqrt{C_{\mathrm{O}}}\approx 1$ for the edge and orbital current respectively. 

\begin{figure}
	\centering
	\includegraphics[width=\linewidth]{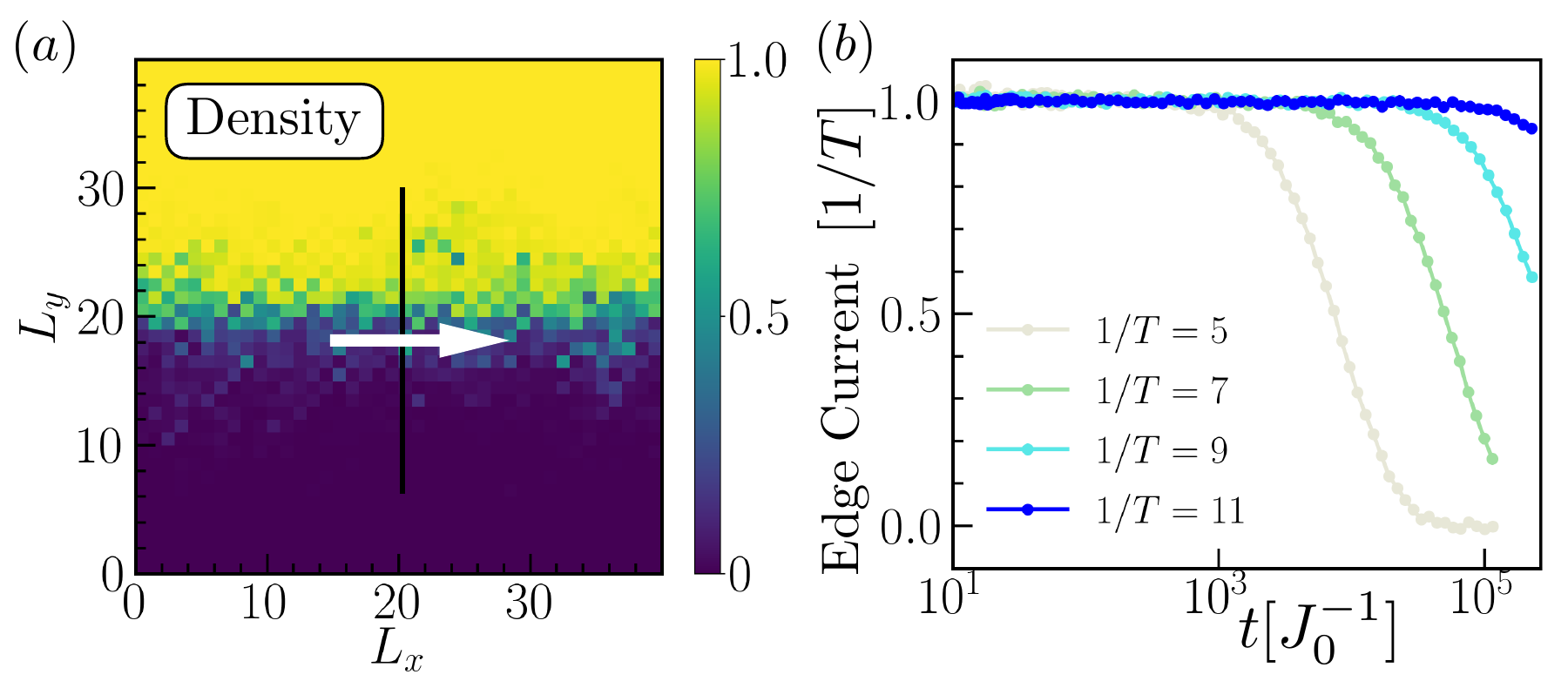}
	\caption{(a) Particle density for a single $3-$RMD realization at time $t\approx400J_0^{-1}$ for $1/T=9$. (b) Edge current averaged over a single block at the boundary between filled and empty sites with $3-$RMD for different driving frequencies. We use parameters $\delta J=1J_0, h_{\mathrm{max}}=10J_0,\delta h=7J_0$. $T^{-1}$ is in units of $J_0$. }
	\label{fig:current_random}
\end{figure}
When switching from quasi-periodic TM driving  to the RMD, the prethermal topological phase remains robust. However, as the recursive quasi-periodic structure is missing~[35], to simulate a sufficiently long time evolution ($t\sim 10^6J_0^{-1}$), the size of the square lattice is limited to $40\times 40$. In Fig.~\ref{fig:current_random}, we illustrate the dynamics for $n=3$ RMD, in the case where the upper half of a cylinder is fully filled as the initial state. In Fig.~\ref{fig:current_random} (a), the particle density at time {$t\approx 400J_0^{-1} $} is depicted. The density only changes significantly in a strip centered around the boundary of the filled region, similarly to Fig.~\ref{fig:current} (a). As shown in Fig.~\ref{fig:current_random} (b), the current across the vertical black cut
remains close to a constant quantized value for a long time, confirming the existence of the prethermal ARMDI. In the SM we further verify the dependence of the prethermal lifetime on the driving rate, which exhibits an algebraic scaling similar to the behavior of the PR as $\tau_{\mathrm{RMD}}\sim T^{-(2n+1)}$.

\section{Discussion and Outlook}
By constructing a concrete example, we have established that aperiodically driven systems can host novel non-equilibrium topological phases of matter without any equilibrium counterparts. The price to pay for relieving the constraint of TTS is that the ARMDI is strictly speaking only transient, disappearing in the asymptotic long-time limit. However, its prethermal lifetime can be tuned arbitrarily long with a controllable universal scaling of the heating times.   

Regarding the experimental feasibility of our proposal, we note that the (periodically-driven) AFAI has recently been realized in cold atom quantum simulators~\cite{wintersperger2020realization}. We expect that the RMD drives proposed here can be naturally implemented in a similar fashion. In that context, recent simulation platforms studying prethermalization, e.g., in trapped ions~\cite{kyprianidis2021observation} or cold atoms~\cite{rubio2020floquet}, would also permit a study of the universal scaling of the prethermal time scale as a function of $n$. 

A fundamental and open question is  whether there exist aperiodically driven topological phases stable for infinitely long times, e.g., in discrete versus continuous driven clean or disordered systems. Finally, the role of interactions for the stability of the ARMDI prethermal phase is a very interesting and challenging problem, which is beyond the reach of numerical methods and is thus an ideal candidate for quantum simulators. 

\section{Acknowledgements}  
HZ acknowledges support from a Doctoral-Program Fellowship of the German Academic Exchange Service (DAAD). We acknowledge support from the Imperial-TUM flagship partnership.
MR gratefully acknowledges the support of the European Research Council (ERC) under the European Union Horizon 2020 Research and Innovation Programme (Grant Agreement No.~678862), and the Villum Foundation. This work was partly supported by the Deutsche Forschungsgemeinschaft under grants SFB 1143 (project-id 247310070) and the cluster of excellence ct.qmat (EXC 2147, project-id 390858490). The research is part of the Munich Quantum Valley, which is supported by the Bavarian state government with funds from the Hightech Agenda Bayern Plus.

\bibliography{Topology} 

\begin{thebibliography}{45}%
\makeatletter
\providecommand \@ifxundefined [1]{%
 \@ifx{#1\undefined}
}%
\providecommand \@ifnum [1]{%
 \ifnum #1\expandafter \@firstoftwo
 \else \expandafter \@secondoftwo
 \fi
}%
\providecommand \@ifx [1]{%
 \ifx #1\expandafter \@firstoftwo
 \else \expandafter \@secondoftwo
 \fi
}%
\providecommand \natexlab [1]{#1}%
\providecommand \enquote  [1]{``#1''}%
\providecommand \bibnamefont  [1]{#1}%
\providecommand \bibfnamefont [1]{#1}%
\providecommand \citenamefont [1]{#1}%
\providecommand \href@noop [0]{\@secondoftwo}%
\providecommand \href [0]{\begingroup \@sanitize@url \@href}%
\providecommand \@href[1]{\@@startlink{#1}\@@href}%
\providecommand \@@href[1]{\endgroup#1\@@endlink}%
\providecommand \@sanitize@url [0]{\catcode `\\12\catcode `\$12\catcode
  `\&12\catcode `\#12\catcode `\^12\catcode `\_12\catcode `\%12\relax}%
\providecommand \@@startlink[1]{}%
\providecommand \@@endlink[0]{}%
\providecommand \url  [0]{\begingroup\@sanitize@url \@url }%
\providecommand \@url [1]{\endgroup\@href {#1}{\urlprefix }}%
\providecommand \urlprefix  [0]{URL }%
\providecommand \Eprint [0]{\href }%
\providecommand \doibase [0]{http://dx.doi.org/}%
\providecommand \selectlanguage [0]{\@gobble}%
\providecommand \bibinfo  [0]{\@secondoftwo}%
\providecommand \bibfield  [0]{\@secondoftwo}%
\providecommand \translation [1]{[#1]}%
\providecommand \BibitemOpen [0]{}%
\providecommand \bibitemStop [0]{}%
\providecommand \bibitemNoStop [0]{.\EOS\space}%
\providecommand \EOS [0]{\spacefactor3000\relax}%
\providecommand \BibitemShut  [1]{\csname bibitem#1\endcsname}%
\let\auto@bib@innerbib\@empty
\bibitem [{\citenamefont {Arute}\ \emph {et~al.}(2019)\citenamefont {Arute},
  \citenamefont {Arya}, \citenamefont {Babbush}, \citenamefont {Bacon},
  \citenamefont {Bardin}, \citenamefont {Barends}, \citenamefont {Biswas},
  \citenamefont {Boixo}, \citenamefont {Brandao}, \citenamefont {Buell} \emph
  {et~al.}}]{arute2019quantum}%
  \BibitemOpen
  \bibfield  {author} {\bibinfo {author} {\bibfnamefont {F.}~\bibnamefont
  {Arute}}, \bibinfo {author} {\bibfnamefont {K.}~\bibnamefont {Arya}},
  \bibinfo {author} {\bibfnamefont {R.}~\bibnamefont {Babbush}}, \bibinfo
  {author} {\bibfnamefont {D.}~\bibnamefont {Bacon}}, \bibinfo {author}
  {\bibfnamefont {J.~C.}\ \bibnamefont {Bardin}}, \bibinfo {author}
  {\bibfnamefont {R.}~\bibnamefont {Barends}}, \bibinfo {author} {\bibfnamefont
  {R.}~\bibnamefont {Biswas}}, \bibinfo {author} {\bibfnamefont
  {S.}~\bibnamefont {Boixo}}, \bibinfo {author} {\bibfnamefont {F.~G.}\
  \bibnamefont {Brandao}}, \bibinfo {author} {\bibfnamefont {D.~A.}\
  \bibnamefont {Buell}},  \emph {et~al.},\ }\href@noop {} {\bibfield  {journal}
  {\bibinfo  {journal} {Nature}\ }\textbf {\bibinfo {volume} {574}},\ \bibinfo
  {pages} {505} (\bibinfo {year} {2019})}\BibitemShut {NoStop}%
\bibitem [{\citenamefont {Gross}\ and\ \citenamefont
  {Bloch}(2017)}]{gross2017quantum}%
  \BibitemOpen
  \bibfield  {author} {\bibinfo {author} {\bibfnamefont {C.}~\bibnamefont
  {Gross}}\ and\ \bibinfo {author} {\bibfnamefont {I.}~\bibnamefont {Bloch}},\
  }\href@noop {} {\bibfield  {journal} {\bibinfo  {journal} {Science}\ }\textbf
  {\bibinfo {volume} {357}},\ \bibinfo {pages} {995} (\bibinfo {year}
  {2017})}\BibitemShut {NoStop}%
\bibitem [{\citenamefont {Martinez}\ \emph {et~al.}(2016)\citenamefont
  {Martinez}, \citenamefont {Muschik}, \citenamefont {Schindler}, \citenamefont
  {Nigg}, \citenamefont {Erhard}, \citenamefont {Heyl}, \citenamefont {Hauke},
  \citenamefont {Dalmonte}, \citenamefont {Monz}, \citenamefont {Zoller} \emph
  {et~al.}}]{martinez2016real}%
  \BibitemOpen
  \bibfield  {author} {\bibinfo {author} {\bibfnamefont {E.~A.}\ \bibnamefont
  {Martinez}}, \bibinfo {author} {\bibfnamefont {C.~A.}\ \bibnamefont
  {Muschik}}, \bibinfo {author} {\bibfnamefont {P.}~\bibnamefont {Schindler}},
  \bibinfo {author} {\bibfnamefont {D.}~\bibnamefont {Nigg}}, \bibinfo {author}
  {\bibfnamefont {A.}~\bibnamefont {Erhard}}, \bibinfo {author} {\bibfnamefont
  {M.}~\bibnamefont {Heyl}}, \bibinfo {author} {\bibfnamefont {P.}~\bibnamefont
  {Hauke}}, \bibinfo {author} {\bibfnamefont {M.}~\bibnamefont {Dalmonte}},
  \bibinfo {author} {\bibfnamefont {T.}~\bibnamefont {Monz}}, \bibinfo {author}
  {\bibfnamefont {P.}~\bibnamefont {Zoller}},  \emph {et~al.},\ }\href@noop {}
  {\bibfield  {journal} {\bibinfo  {journal} {Nature}\ }\textbf {\bibinfo
  {volume} {534}},\ \bibinfo {pages} {516} (\bibinfo {year}
  {2016})}\BibitemShut {NoStop}%
\bibitem [{\citenamefont {Maczewsky}\ \emph {et~al.}(2017)\citenamefont
  {Maczewsky}, \citenamefont {Zeuner}, \citenamefont {Nolte},\ and\
  \citenamefont {Szameit}}]{maczewsky2017observation}%
  \BibitemOpen
  \bibfield  {author} {\bibinfo {author} {\bibfnamefont {L.~J.}\ \bibnamefont
  {Maczewsky}}, \bibinfo {author} {\bibfnamefont {J.~M.}\ \bibnamefont
  {Zeuner}}, \bibinfo {author} {\bibfnamefont {S.}~\bibnamefont {Nolte}}, \
  and\ \bibinfo {author} {\bibfnamefont {A.}~\bibnamefont {Szameit}},\
  }\href@noop {} {\bibfield  {journal} {\bibinfo  {journal} {Nature
  communications}\ }\textbf {\bibinfo {volume} {8}},\ \bibinfo {pages} {1}
  (\bibinfo {year} {2017})}\BibitemShut {NoStop}%
\bibitem [{\citenamefont {Yang}\ \emph {et~al.}(2020)\citenamefont {Yang},
  \citenamefont {Sun}, \citenamefont {Ott}, \citenamefont {Wang}, \citenamefont
  {Zache}, \citenamefont {Halimeh}, \citenamefont {Yuan}, \citenamefont
  {Hauke},\ and\ \citenamefont {Pan}}]{yang2020observation}%
  \BibitemOpen
  \bibfield  {author} {\bibinfo {author} {\bibfnamefont {B.}~\bibnamefont
  {Yang}}, \bibinfo {author} {\bibfnamefont {H.}~\bibnamefont {Sun}}, \bibinfo
  {author} {\bibfnamefont {R.}~\bibnamefont {Ott}}, \bibinfo {author}
  {\bibfnamefont {H.-Y.}\ \bibnamefont {Wang}}, \bibinfo {author}
  {\bibfnamefont {T.~V.}\ \bibnamefont {Zache}}, \bibinfo {author}
  {\bibfnamefont {J.~C.}\ \bibnamefont {Halimeh}}, \bibinfo {author}
  {\bibfnamefont {Z.-S.}\ \bibnamefont {Yuan}}, \bibinfo {author}
  {\bibfnamefont {P.}~\bibnamefont {Hauke}}, \ and\ \bibinfo {author}
  {\bibfnamefont {J.-W.}\ \bibnamefont {Pan}},\ }\href@noop {} {\bibfield
  {journal} {\bibinfo  {journal} {Nature}\ }\textbf {\bibinfo {volume} {587}},\
  \bibinfo {pages} {392} (\bibinfo {year} {2020})}\BibitemShut {NoStop}%
\bibitem [{\citenamefont {Schweizer}\ \emph {et~al.}(2019)\citenamefont
  {Schweizer}, \citenamefont {Grusdt}, \citenamefont {Berngruber},
  \citenamefont {Barbiero}, \citenamefont {Demler}, \citenamefont {Goldman},
  \citenamefont {Bloch},\ and\ \citenamefont
  {Aidelsburger}}]{schweizer2019floquet}%
  \BibitemOpen
  \bibfield  {author} {\bibinfo {author} {\bibfnamefont {C.}~\bibnamefont
  {Schweizer}}, \bibinfo {author} {\bibfnamefont {F.}~\bibnamefont {Grusdt}},
  \bibinfo {author} {\bibfnamefont {M.}~\bibnamefont {Berngruber}}, \bibinfo
  {author} {\bibfnamefont {L.}~\bibnamefont {Barbiero}}, \bibinfo {author}
  {\bibfnamefont {E.}~\bibnamefont {Demler}}, \bibinfo {author} {\bibfnamefont
  {N.}~\bibnamefont {Goldman}}, \bibinfo {author} {\bibfnamefont
  {I.}~\bibnamefont {Bloch}}, \ and\ \bibinfo {author} {\bibfnamefont
  {M.}~\bibnamefont {Aidelsburger}},\ }\href@noop {} {\bibfield  {journal}
  {\bibinfo  {journal} {Nature Physics}\ }\textbf {\bibinfo {volume} {15}},\
  \bibinfo {pages} {1168} (\bibinfo {year} {2019})}\BibitemShut {NoStop}%
\bibitem [{\citenamefont {Oka}\ and\ \citenamefont
  {Kitamura}(2019)}]{oka2019floquet}%
  \BibitemOpen
  \bibfield  {author} {\bibinfo {author} {\bibfnamefont {T.}~\bibnamefont
  {Oka}}\ and\ \bibinfo {author} {\bibfnamefont {S.}~\bibnamefont {Kitamura}},\
  }\href@noop {} {\bibfield  {journal} {\bibinfo  {journal} {Annual Review of
  Condensed Matter Physics}\ }\textbf {\bibinfo {volume} {10}},\ \bibinfo
  {pages} {387} (\bibinfo {year} {2019})}\BibitemShut {NoStop}%
\bibitem [{\citenamefont {Kitagawa}\ \emph {et~al.}(2010)\citenamefont
  {Kitagawa}, \citenamefont {Berg}, \citenamefont {Rudner},\ and\ \citenamefont
  {Demler}}]{kitagawa2010topological}%
  \BibitemOpen
  \bibfield  {author} {\bibinfo {author} {\bibfnamefont {T.}~\bibnamefont
  {Kitagawa}}, \bibinfo {author} {\bibfnamefont {E.}~\bibnamefont {Berg}},
  \bibinfo {author} {\bibfnamefont {M.}~\bibnamefont {Rudner}}, \ and\ \bibinfo
  {author} {\bibfnamefont {E.}~\bibnamefont {Demler}},\ }\href@noop {}
  {\bibfield  {journal} {\bibinfo  {journal} {Physical Review B}\ }\textbf
  {\bibinfo {volume} {82}},\ \bibinfo {pages} {235114} (\bibinfo {year}
  {2010})}\BibitemShut {NoStop}%
\bibitem [{\citenamefont {Khemani}\ \emph {et~al.}(2016)\citenamefont
  {Khemani}, \citenamefont {Lazarides}, \citenamefont {Moessner},\ and\
  \citenamefont {Sondhi}}]{khemani2016phase}%
  \BibitemOpen
  \bibfield  {author} {\bibinfo {author} {\bibfnamefont {V.}~\bibnamefont
  {Khemani}}, \bibinfo {author} {\bibfnamefont {A.}~\bibnamefont {Lazarides}},
  \bibinfo {author} {\bibfnamefont {R.}~\bibnamefont {Moessner}}, \ and\
  \bibinfo {author} {\bibfnamefont {S.~L.}\ \bibnamefont {Sondhi}},\
  }\href@noop {} {\bibfield  {journal} {\bibinfo  {journal} {Physical review
  letters}\ }\textbf {\bibinfo {volume} {116}},\ \bibinfo {pages} {250401}
  (\bibinfo {year} {2016})}\BibitemShut {NoStop}%
\bibitem [{\citenamefont {Parameswaran}\ and\ \citenamefont
  {Vasseur}(2018)}]{parameswaran2018many}%
  \BibitemOpen
  \bibfield  {author} {\bibinfo {author} {\bibfnamefont {S.}~\bibnamefont
  {Parameswaran}}\ and\ \bibinfo {author} {\bibfnamefont {R.}~\bibnamefont
  {Vasseur}},\ }\href@noop {} {\bibfield  {journal} {\bibinfo  {journal}
  {Reports on Progress in Physics}\ }\textbf {\bibinfo {volume} {81}},\
  \bibinfo {pages} {082501} (\bibinfo {year} {2018})}\BibitemShut {NoStop}%
\bibitem [{\citenamefont {Moessner}\ and\ \citenamefont
  {Sondhi}(2017)}]{moessner2017equilibration}%
  \BibitemOpen
  \bibfield  {author} {\bibinfo {author} {\bibfnamefont {R.}~\bibnamefont
  {Moessner}}\ and\ \bibinfo {author} {\bibfnamefont {S.~L.}\ \bibnamefont
  {Sondhi}},\ }\href@noop {} {\bibfield  {journal} {\bibinfo  {journal} {Nature
  Physics}\ }\textbf {\bibinfo {volume} {13}},\ \bibinfo {pages} {424}
  (\bibinfo {year} {2017})}\BibitemShut {NoStop}%
\bibitem [{\citenamefont {Else}\ \emph
  {et~al.}(2020{\natexlab{a}})\citenamefont {Else}, \citenamefont {Monroe},
  \citenamefont {Nayak},\ and\ \citenamefont {Yao}}]{else2020discrete}%
  \BibitemOpen
  \bibfield  {author} {\bibinfo {author} {\bibfnamefont {D.~V.}\ \bibnamefont
  {Else}}, \bibinfo {author} {\bibfnamefont {C.}~\bibnamefont {Monroe}},
  \bibinfo {author} {\bibfnamefont {C.}~\bibnamefont {Nayak}}, \ and\ \bibinfo
  {author} {\bibfnamefont {N.~Y.}\ \bibnamefont {Yao}},\ }\href@noop {}
  {\bibfield  {journal} {\bibinfo  {journal} {Annual Review of Condensed Matter
  Physics}\ }\textbf {\bibinfo {volume} {11}},\ \bibinfo {pages} {467}
  (\bibinfo {year} {2020}{\natexlab{a}})}\BibitemShut {NoStop}%
\bibitem [{\citenamefont {Khemani}\ \emph {et~al.}(2019)\citenamefont
  {Khemani}, \citenamefont {Moessner},\ and\ \citenamefont
  {Sondhi}}]{khemani2019brief}%
  \BibitemOpen
  \bibfield  {author} {\bibinfo {author} {\bibfnamefont {V.}~\bibnamefont
  {Khemani}}, \bibinfo {author} {\bibfnamefont {R.}~\bibnamefont {Moessner}}, \
  and\ \bibinfo {author} {\bibfnamefont {S.}~\bibnamefont {Sondhi}},\
  }\href@noop {} {\bibfield  {journal} {\bibinfo  {journal} {arXiv preprint
  arXiv:1910.10745}\ } (\bibinfo {year} {2019})}\BibitemShut {NoStop}%
\bibitem [{\citenamefont {Pizzi}\ \emph
  {et~al.}(2021{\natexlab{a}})\citenamefont {Pizzi}, \citenamefont {Knolle},\
  and\ \citenamefont {Nunnenkamp}}]{pizzi2021higher}%
  \BibitemOpen
  \bibfield  {author} {\bibinfo {author} {\bibfnamefont {A.}~\bibnamefont
  {Pizzi}}, \bibinfo {author} {\bibfnamefont {J.}~\bibnamefont {Knolle}}, \
  and\ \bibinfo {author} {\bibfnamefont {A.}~\bibnamefont {Nunnenkamp}},\
  }\href@noop {} {\bibfield  {journal} {\bibinfo  {journal} {Nature
  communications}\ }\textbf {\bibinfo {volume} {12}},\ \bibinfo {pages} {1}
  (\bibinfo {year} {2021}{\natexlab{a}})}\BibitemShut {NoStop}%
\bibitem [{\citenamefont {Titum}\ \emph {et~al.}(2016)\citenamefont {Titum},
  \citenamefont {Berg}, \citenamefont {Rudner}, \citenamefont {Refael},\ and\
  \citenamefont {Lindner}}]{titum2016anomalous}%
  \BibitemOpen
  \bibfield  {author} {\bibinfo {author} {\bibfnamefont {P.}~\bibnamefont
  {Titum}}, \bibinfo {author} {\bibfnamefont {E.}~\bibnamefont {Berg}},
  \bibinfo {author} {\bibfnamefont {M.~S.}\ \bibnamefont {Rudner}}, \bibinfo
  {author} {\bibfnamefont {G.}~\bibnamefont {Refael}}, \ and\ \bibinfo {author}
  {\bibfnamefont {N.~H.}\ \bibnamefont {Lindner}},\ }\href@noop {} {\bibfield
  {journal} {\bibinfo  {journal} {Physical Review X}\ }\textbf {\bibinfo
  {volume} {6}},\ \bibinfo {pages} {021013} (\bibinfo {year}
  {2016})}\BibitemShut {NoStop}%
\bibitem [{\citenamefont {Rudner}\ \emph {et~al.}(2013)\citenamefont {Rudner},
  \citenamefont {Lindner}, \citenamefont {Berg},\ and\ \citenamefont
  {Levin}}]{rudner2013anomalous}%
  \BibitemOpen
  \bibfield  {author} {\bibinfo {author} {\bibfnamefont {M.~S.}\ \bibnamefont
  {Rudner}}, \bibinfo {author} {\bibfnamefont {N.~H.}\ \bibnamefont {Lindner}},
  \bibinfo {author} {\bibfnamefont {E.}~\bibnamefont {Berg}}, \ and\ \bibinfo
  {author} {\bibfnamefont {M.}~\bibnamefont {Levin}},\ }\href@noop {}
  {\bibfield  {journal} {\bibinfo  {journal} {Physical Review X}\ }\textbf
  {\bibinfo {volume} {3}},\ \bibinfo {pages} {031005} (\bibinfo {year}
  {2013})}\BibitemShut {NoStop}%
\bibitem [{\citenamefont {Nathan}\ \emph {et~al.}(2019)\citenamefont {Nathan},
  \citenamefont {Abanin}, \citenamefont {Berg}, \citenamefont {Lindner},\ and\
  \citenamefont {Rudner}}]{nathan2019anomalous}%
  \BibitemOpen
  \bibfield  {author} {\bibinfo {author} {\bibfnamefont {F.}~\bibnamefont
  {Nathan}}, \bibinfo {author} {\bibfnamefont {D.}~\bibnamefont {Abanin}},
  \bibinfo {author} {\bibfnamefont {E.}~\bibnamefont {Berg}}, \bibinfo {author}
  {\bibfnamefont {N.~H.}\ \bibnamefont {Lindner}}, \ and\ \bibinfo {author}
  {\bibfnamefont {M.~S.}\ \bibnamefont {Rudner}},\ }\href@noop {} {\bibfield
  {journal} {\bibinfo  {journal} {Physical Review B}\ }\textbf {\bibinfo
  {volume} {99}},\ \bibinfo {pages} {195133} (\bibinfo {year}
  {2019})}\BibitemShut {NoStop}%
\bibitem [{\citenamefont {Dumitrescu}\ \emph {et~al.}(2018)\citenamefont
  {Dumitrescu}, \citenamefont {Vasseur},\ and\ \citenamefont
  {Potter}}]{dumitrescu2018logarithmically}%
  \BibitemOpen
  \bibfield  {author} {\bibinfo {author} {\bibfnamefont {P.~T.}\ \bibnamefont
  {Dumitrescu}}, \bibinfo {author} {\bibfnamefont {R.}~\bibnamefont {Vasseur}},
  \ and\ \bibinfo {author} {\bibfnamefont {A.~C.}\ \bibnamefont {Potter}},\
  }\href@noop {} {\bibfield  {journal} {\bibinfo  {journal} {Physical review
  letters}\ }\textbf {\bibinfo {volume} {120}},\ \bibinfo {pages} {070602}
  (\bibinfo {year} {2018})}\BibitemShut {NoStop}%
\bibitem [{\citenamefont {Friedman}\ \emph {et~al.}(2020)\citenamefont
  {Friedman}, \citenamefont {Ware}, \citenamefont {Vasseur},\ and\
  \citenamefont {Potter}}]{friedman2020topological}%
  \BibitemOpen
  \bibfield  {author} {\bibinfo {author} {\bibfnamefont {A.~J.}\ \bibnamefont
  {Friedman}}, \bibinfo {author} {\bibfnamefont {B.}~\bibnamefont {Ware}},
  \bibinfo {author} {\bibfnamefont {R.}~\bibnamefont {Vasseur}}, \ and\
  \bibinfo {author} {\bibfnamefont {A.~C.}\ \bibnamefont {Potter}},\
  }\href@noop {} {\bibfield  {journal} {\bibinfo  {journal} {arXiv preprint
  arXiv:2009.03314}\ } (\bibinfo {year} {2020})}\BibitemShut {NoStop}%
\bibitem [{\citenamefont {Zhao}\ \emph {et~al.}(2019)\citenamefont {Zhao},
  \citenamefont {Mintert},\ and\ \citenamefont {Knolle}}]{zhao2019floquet}%
  \BibitemOpen
  \bibfield  {author} {\bibinfo {author} {\bibfnamefont {H.}~\bibnamefont
  {Zhao}}, \bibinfo {author} {\bibfnamefont {F.}~\bibnamefont {Mintert}}, \
  and\ \bibinfo {author} {\bibfnamefont {J.}~\bibnamefont {Knolle}},\
  }\href@noop {} {\bibfield  {journal} {\bibinfo  {journal} {Physical Review
  B}\ }\textbf {\bibinfo {volume} {100}},\ \bibinfo {pages} {134302} (\bibinfo
  {year} {2019})}\BibitemShut {NoStop}%
\bibitem [{\citenamefont {Else}\ \emph
  {et~al.}(2020{\natexlab{b}})\citenamefont {Else}, \citenamefont {Ho},\ and\
  \citenamefont {Dumitrescu}}]{else2020long}%
  \BibitemOpen
  \bibfield  {author} {\bibinfo {author} {\bibfnamefont {D.~V.}\ \bibnamefont
  {Else}}, \bibinfo {author} {\bibfnamefont {W.~W.}\ \bibnamefont {Ho}}, \ and\
  \bibinfo {author} {\bibfnamefont {P.~T.}\ \bibnamefont {Dumitrescu}},\
  }\href@noop {} {\bibfield  {journal} {\bibinfo  {journal} {Physical Review
  X}\ }\textbf {\bibinfo {volume} {10}},\ \bibinfo {pages} {021032} (\bibinfo
  {year} {2020}{\natexlab{b}})}\BibitemShut {NoStop}%
\bibitem [{\citenamefont {Long}\ \emph {et~al.}(2021)\citenamefont {Long},
  \citenamefont {Crowley},\ and\ \citenamefont {Chandran}}]{long2021many}%
  \BibitemOpen
  \bibfield  {author} {\bibinfo {author} {\bibfnamefont {D.~M.}\ \bibnamefont
  {Long}}, \bibinfo {author} {\bibfnamefont {P.~J.}\ \bibnamefont {Crowley}}, \
  and\ \bibinfo {author} {\bibfnamefont {A.}~\bibnamefont {Chandran}},\
  }\href@noop {} {\bibfield  {journal} {\bibinfo  {journal} {arXiv preprint
  arXiv:2108.04834}\ } (\bibinfo {year} {2021})}\BibitemShut {NoStop}%
\bibitem [{\citenamefont {Martin}\ \emph {et~al.}(2017)\citenamefont {Martin},
  \citenamefont {Refael},\ and\ \citenamefont
  {Halperin}}]{martin2017topological}%
  \BibitemOpen
  \bibfield  {author} {\bibinfo {author} {\bibfnamefont {I.}~\bibnamefont
  {Martin}}, \bibinfo {author} {\bibfnamefont {G.}~\bibnamefont {Refael}}, \
  and\ \bibinfo {author} {\bibfnamefont {B.}~\bibnamefont {Halperin}},\
  }\href@noop {} {\bibfield  {journal} {\bibinfo  {journal} {Physical Review
  X}\ }\textbf {\bibinfo {volume} {7}},\ \bibinfo {pages} {041008} (\bibinfo
  {year} {2017})}\BibitemShut {NoStop}%
\bibitem [{\citenamefont {Nathan}\ \emph {et~al.}(2020)\citenamefont {Nathan},
  \citenamefont {Ge}, \citenamefont {Gazit}, \citenamefont {Rudner},\ and\
  \citenamefont {Kolodrubetz}}]{nathan2020quasiperiodic}%
  \BibitemOpen
  \bibfield  {author} {\bibinfo {author} {\bibfnamefont {F.}~\bibnamefont
  {Nathan}}, \bibinfo {author} {\bibfnamefont {R.}~\bibnamefont {Ge}}, \bibinfo
  {author} {\bibfnamefont {S.}~\bibnamefont {Gazit}}, \bibinfo {author}
  {\bibfnamefont {M.~S.}\ \bibnamefont {Rudner}}, \ and\ \bibinfo {author}
  {\bibfnamefont {M.}~\bibnamefont {Kolodrubetz}},\ }\href@noop {} {\bibfield
  {journal} {\bibinfo  {journal} {arXiv preprint arXiv:2010.11485}\ } (\bibinfo
  {year} {2020})}\BibitemShut {NoStop}%
\bibitem [{\citenamefont {Rieder}\ \emph {et~al.}(2018)\citenamefont {Rieder},
  \citenamefont {Sieberer}, \citenamefont {Fischer},\ and\ \citenamefont
  {Fulga}}]{rieder2018localization}%
  \BibitemOpen
  \bibfield  {author} {\bibinfo {author} {\bibfnamefont {M.-T.}\ \bibnamefont
  {Rieder}}, \bibinfo {author} {\bibfnamefont {L.~M.}\ \bibnamefont
  {Sieberer}}, \bibinfo {author} {\bibfnamefont {M.~H.}\ \bibnamefont
  {Fischer}}, \ and\ \bibinfo {author} {\bibfnamefont {I.~C.}\ \bibnamefont
  {Fulga}},\ }\href@noop {} {\bibfield  {journal} {\bibinfo  {journal}
  {Physical review letters}\ }\textbf {\bibinfo {volume} {120}},\ \bibinfo
  {pages} {216801} (\bibinfo {year} {2018})}\BibitemShut {NoStop}%
\bibitem [{\citenamefont {Timms}\ and\ \citenamefont
  {Kolodrubetz}(2020)}]{timms2020quantized}%
  \BibitemOpen
  \bibfield  {author} {\bibinfo {author} {\bibfnamefont {C.~I.}\ \bibnamefont
  {Timms}}\ and\ \bibinfo {author} {\bibfnamefont {M.~H.}\ \bibnamefont
  {Kolodrubetz}},\ }\href@noop {} {\bibfield  {journal} {\bibinfo  {journal}
  {arXiv preprint arXiv:2006.10736}\ } (\bibinfo {year} {2020})}\BibitemShut
  {NoStop}%
\bibitem [{\citenamefont {Crowley}\ \emph {et~al.}(2019)\citenamefont
  {Crowley}, \citenamefont {Martin},\ and\ \citenamefont
  {Chandran}}]{crowley2019topological}%
  \BibitemOpen
  \bibfield  {author} {\bibinfo {author} {\bibfnamefont {P.~J.}\ \bibnamefont
  {Crowley}}, \bibinfo {author} {\bibfnamefont {I.}~\bibnamefont {Martin}}, \
  and\ \bibinfo {author} {\bibfnamefont {A.}~\bibnamefont {Chandran}},\
  }\href@noop {} {\bibfield  {journal} {\bibinfo  {journal} {Physical Review
  B}\ }\textbf {\bibinfo {volume} {99}},\ \bibinfo {pages} {064306} (\bibinfo
  {year} {2019})}\BibitemShut {NoStop}%
\bibitem [{\citenamefont {Zhao}\ \emph
  {et~al.}(2021{\natexlab{a}})\citenamefont {Zhao}, \citenamefont {Mintert},
  \citenamefont {Knolle},\ and\ \citenamefont
  {Moessner}}]{zhao2021localization}%
  \BibitemOpen
  \bibfield  {author} {\bibinfo {author} {\bibfnamefont {H.}~\bibnamefont
  {Zhao}}, \bibinfo {author} {\bibfnamefont {F.}~\bibnamefont {Mintert}},
  \bibinfo {author} {\bibfnamefont {J.}~\bibnamefont {Knolle}}, \ and\ \bibinfo
  {author} {\bibfnamefont {R.}~\bibnamefont {Moessner}},\ }\href@noop {}
  {\bibfield  {journal} {\bibinfo  {journal} {arXiv preprint arXiv:2111.13558}\
  } (\bibinfo {year} {2021}{\natexlab{a}})}\BibitemShut {NoStop}%
\bibitem [{\citenamefont {Zhao}\ \emph
  {et~al.}(2021{\natexlab{b}})\citenamefont {Zhao}, \citenamefont {Mintert},
  \citenamefont {Moessner},\ and\ \citenamefont {Knolle}}]{zhao2021random}%
  \BibitemOpen
  \bibfield  {author} {\bibinfo {author} {\bibfnamefont {H.}~\bibnamefont
  {Zhao}}, \bibinfo {author} {\bibfnamefont {F.}~\bibnamefont {Mintert}},
  \bibinfo {author} {\bibfnamefont {R.}~\bibnamefont {Moessner}}, \ and\
  \bibinfo {author} {\bibfnamefont {J.}~\bibnamefont {Knolle}},\ }\href@noop {}
  {\bibfield  {journal} {\bibinfo  {journal} {Physical Review Letters}\
  }\textbf {\bibinfo {volume} {126}},\ \bibinfo {pages} {040601} (\bibinfo
  {year} {2021}{\natexlab{b}})}\BibitemShut {NoStop}%
\bibitem [{\citenamefont {Mori}\ \emph {et~al.}(2021)\citenamefont {Mori},
  \citenamefont {Zhao}, \citenamefont {Mintert}, \citenamefont {Knolle},\ and\
  \citenamefont {Moessner}}]{mori2021rigorous}%
  \BibitemOpen
  \bibfield  {author} {\bibinfo {author} {\bibfnamefont {T.}~\bibnamefont
  {Mori}}, \bibinfo {author} {\bibfnamefont {H.}~\bibnamefont {Zhao}}, \bibinfo
  {author} {\bibfnamefont {F.}~\bibnamefont {Mintert}}, \bibinfo {author}
  {\bibfnamefont {J.}~\bibnamefont {Knolle}}, \ and\ \bibinfo {author}
  {\bibfnamefont {R.}~\bibnamefont {Moessner}},\ }\href@noop {} {\bibfield
  {journal} {\bibinfo  {journal} {arXiv preprint arXiv:2101.07065}\ } (\bibinfo
  {year} {2021})}\BibitemShut {NoStop}%
\bibitem [{\citenamefont {Nandy}\ \emph {et~al.}(2017)\citenamefont {Nandy},
  \citenamefont {Sen},\ and\ \citenamefont {Sen}}]{nandy2017aperiodically}%
  \BibitemOpen
  \bibfield  {author} {\bibinfo {author} {\bibfnamefont {S.}~\bibnamefont
  {Nandy}}, \bibinfo {author} {\bibfnamefont {A.}~\bibnamefont {Sen}}, \ and\
  \bibinfo {author} {\bibfnamefont {D.}~\bibnamefont {Sen}},\ }\href@noop {}
  {\bibfield  {journal} {\bibinfo  {journal} {Physical Review X}\ }\textbf
  {\bibinfo {volume} {7}},\ \bibinfo {pages} {031034} (\bibinfo {year}
  {2017})}\BibitemShut {NoStop}%
\bibitem [{\citenamefont {Kuwahara}\ \emph {et~al.}(2016)\citenamefont
  {Kuwahara}, \citenamefont {Mori},\ and\ \citenamefont
  {Saito}}]{kuwahara2016floquet}%
  \BibitemOpen
  \bibfield  {author} {\bibinfo {author} {\bibfnamefont {T.}~\bibnamefont
  {Kuwahara}}, \bibinfo {author} {\bibfnamefont {T.}~\bibnamefont {Mori}}, \
  and\ \bibinfo {author} {\bibfnamefont {K.}~\bibnamefont {Saito}},\
  }\href@noop {} {\bibfield  {journal} {\bibinfo  {journal} {Annals of
  Physics}\ }\textbf {\bibinfo {volume} {367}},\ \bibinfo {pages} {96}
  (\bibinfo {year} {2016})}\BibitemShut {NoStop}%
\bibitem [{\citenamefont {Abanin}\ \emph {et~al.}(2017)\citenamefont {Abanin},
  \citenamefont {De~Roeck}, \citenamefont {Ho},\ and\ \citenamefont
  {Huveneers}}]{abanin2017rigorous}%
  \BibitemOpen
  \bibfield  {author} {\bibinfo {author} {\bibfnamefont {D.}~\bibnamefont
  {Abanin}}, \bibinfo {author} {\bibfnamefont {W.}~\bibnamefont {De~Roeck}},
  \bibinfo {author} {\bibfnamefont {W.~W.}\ \bibnamefont {Ho}}, \ and\ \bibinfo
  {author} {\bibfnamefont {F.}~\bibnamefont {Huveneers}},\ }\href@noop {}
  {\bibfield  {journal} {\bibinfo  {journal} {Communications in Mathematical
  Physics}\ }\textbf {\bibinfo {volume} {354}},\ \bibinfo {pages} {809}
  (\bibinfo {year} {2017})}\BibitemShut {NoStop}%
\bibitem [{\citenamefont {Machado}\ \emph {et~al.}(2020)\citenamefont
  {Machado}, \citenamefont {Else}, \citenamefont {Kahanamoku-Meyer},
  \citenamefont {Nayak},\ and\ \citenamefont {Yao}}]{machado2020long}%
  \BibitemOpen
  \bibfield  {author} {\bibinfo {author} {\bibfnamefont {F.}~\bibnamefont
  {Machado}}, \bibinfo {author} {\bibfnamefont {D.~V.}\ \bibnamefont {Else}},
  \bibinfo {author} {\bibfnamefont {G.~D.}\ \bibnamefont {Kahanamoku-Meyer}},
  \bibinfo {author} {\bibfnamefont {C.}~\bibnamefont {Nayak}}, \ and\ \bibinfo
  {author} {\bibfnamefont {N.~Y.}\ \bibnamefont {Yao}},\ }\href@noop {}
  {\bibfield  {journal} {\bibinfo  {journal} {Physical Review X}\ }\textbf
  {\bibinfo {volume} {10}},\ \bibinfo {pages} {011043} (\bibinfo {year}
  {2020})}\BibitemShut {NoStop}%
\bibitem [{\citenamefont {Pizzi}\ \emph
  {et~al.}(2021{\natexlab{b}})\citenamefont {Pizzi}, \citenamefont
  {Nunnenkamp},\ and\ \citenamefont {Knolle}}]{pizzi2021classical}%
  \BibitemOpen
  \bibfield  {author} {\bibinfo {author} {\bibfnamefont {A.}~\bibnamefont
  {Pizzi}}, \bibinfo {author} {\bibfnamefont {A.}~\bibnamefont {Nunnenkamp}}, \
  and\ \bibinfo {author} {\bibfnamefont {J.}~\bibnamefont {Knolle}},\
  }\href@noop {} {\bibfield  {journal} {\bibinfo  {journal} {arXiv preprint
  arXiv:2104.13928}\ } (\bibinfo {year} {2021}{\natexlab{b}})}\BibitemShut
  {NoStop}%
\bibitem [{\citenamefont {Nathan}\ \emph {et~al.}(2017)\citenamefont {Nathan},
  \citenamefont {Rudner}, \citenamefont {Lindner}, \citenamefont {Berg},\ and\
  \citenamefont {Refael}}]{nathan2017quantized}%
  \BibitemOpen
  \bibfield  {author} {\bibinfo {author} {\bibfnamefont {F.}~\bibnamefont
  {Nathan}}, \bibinfo {author} {\bibfnamefont {M.~S.}\ \bibnamefont {Rudner}},
  \bibinfo {author} {\bibfnamefont {N.~H.}\ \bibnamefont {Lindner}}, \bibinfo
  {author} {\bibfnamefont {E.}~\bibnamefont {Berg}}, \ and\ \bibinfo {author}
  {\bibfnamefont {G.}~\bibnamefont {Refael}},\ }\href@noop {} {\bibfield
  {journal} {\bibinfo  {journal} {Physical review letters}\ }\textbf {\bibinfo
  {volume} {119}},\ \bibinfo {pages} {186801} (\bibinfo {year}
  {2017})}\BibitemShut {NoStop}%
\bibitem [{Note1()}]{Note1}%
  \BibitemOpen
  \bibinfo {note} {\label {TM} An important feature of the TM protocol is that,
  due to its recursive nature, only a linearly increasing number of matrix
  multiplications is required to obtain the result of an exponentially long
  time evolution.}\BibitemShut {Stop}%
\bibitem [{Note2()}]{Note2}%
  \BibitemOpen
  \bibinfo {note} {The fundamental block duration for $n$-RMD or TM driving is
  defined as the common duration of the evolution blocks $U^\pm _0$, which we
  assume are equal in length. The lifetime $\tau $ of the prethermal state
  scales algebraically for $n-$RMD as $\tau _{\protect \mathrm {RMD}}\sim
  T_{\protect \mathrm {RMD}}^{-(2n+1)}$, while for TM driving the lifetime
  scales as $\tau _{\protect \mathrm {TM}} \sim e^{C[\protect \qopname \relax
  o{ln}(T_{\protect \mathrm {TM}}^{-1}/g)]^{2}}$ with a constant $C$ and a
  typical local energy scale $g$~~\cite
  {zhao2021random,mori2021rigorous}.}\BibitemShut {Stop}%
\bibitem [{\citenamefont {Janssen}(1998)}]{janssen1998statistics}%
  \BibitemOpen
  \bibfield  {author} {\bibinfo {author} {\bibfnamefont {M.}~\bibnamefont
  {Janssen}},\ }\href@noop {} {\bibfield  {journal} {\bibinfo  {journal}
  {Physics Reports}\ }\textbf {\bibinfo {volume} {295}},\ \bibinfo {pages} {1}
  (\bibinfo {year} {1998})}\BibitemShut {NoStop}%
\bibitem [{Note3()}]{Note3}%
  \BibitemOpen
  \bibinfo {note} {The scaling behavior does not show qualitative dependence on
  the precise values $x$ of the threshold, as long as the state is in the
  prethermal regime at the corresponding time $t_x$.}\BibitemShut {Stop}%
\bibitem [{Note4()}]{Note4}%
  \BibitemOpen
  \bibinfo {note} {Numerical noise can be suppressed in terms of an average
  over the threshold values $0.6/T,0.6\pm 0.3/T,0.6\pm 0.15/T$}\BibitemShut
  {NoStop}%
\bibitem [{\citenamefont {Wintersperger}\ \emph {et~al.}(2020)\citenamefont
  {Wintersperger}, \citenamefont {Braun}, \citenamefont {{\"U}nal},
  \citenamefont {Eckardt}, \citenamefont {Di~Liberto}, \citenamefont {Goldman},
  \citenamefont {Bloch},\ and\ \citenamefont
  {Aidelsburger}}]{wintersperger2020realization}%
  \BibitemOpen
  \bibfield  {author} {\bibinfo {author} {\bibfnamefont {K.}~\bibnamefont
  {Wintersperger}}, \bibinfo {author} {\bibfnamefont {C.}~\bibnamefont
  {Braun}}, \bibinfo {author} {\bibfnamefont {F.~N.}\ \bibnamefont {{\"U}nal}},
  \bibinfo {author} {\bibfnamefont {A.}~\bibnamefont {Eckardt}}, \bibinfo
  {author} {\bibfnamefont {M.}~\bibnamefont {Di~Liberto}}, \bibinfo {author}
  {\bibfnamefont {N.}~\bibnamefont {Goldman}}, \bibinfo {author} {\bibfnamefont
  {I.}~\bibnamefont {Bloch}}, \ and\ \bibinfo {author} {\bibfnamefont
  {M.}~\bibnamefont {Aidelsburger}},\ }\href@noop {} {\bibfield  {journal}
  {\bibinfo  {journal} {Nature Physics}\ }\textbf {\bibinfo {volume} {16}},\
  \bibinfo {pages} {1058} (\bibinfo {year} {2020})}\BibitemShut {NoStop}%
\bibitem [{\citenamefont {Kyprianidis}\ \emph {et~al.}(2021)\citenamefont
  {Kyprianidis}, \citenamefont {Machado}, \citenamefont {Morong}, \citenamefont
  {Becker}, \citenamefont {Collins}, \citenamefont {Else}, \citenamefont
  {Feng}, \citenamefont {Hess}, \citenamefont {Nayak}, \citenamefont {Pagano}
  \emph {et~al.}}]{kyprianidis2021observation}%
  \BibitemOpen
  \bibfield  {author} {\bibinfo {author} {\bibfnamefont {A.}~\bibnamefont
  {Kyprianidis}}, \bibinfo {author} {\bibfnamefont {F.}~\bibnamefont
  {Machado}}, \bibinfo {author} {\bibfnamefont {W.}~\bibnamefont {Morong}},
  \bibinfo {author} {\bibfnamefont {P.}~\bibnamefont {Becker}}, \bibinfo
  {author} {\bibfnamefont {K.~S.}\ \bibnamefont {Collins}}, \bibinfo {author}
  {\bibfnamefont {D.~V.}\ \bibnamefont {Else}}, \bibinfo {author}
  {\bibfnamefont {L.}~\bibnamefont {Feng}}, \bibinfo {author} {\bibfnamefont
  {P.~W.}\ \bibnamefont {Hess}}, \bibinfo {author} {\bibfnamefont
  {C.}~\bibnamefont {Nayak}}, \bibinfo {author} {\bibfnamefont
  {G.}~\bibnamefont {Pagano}},  \emph {et~al.},\ }\href@noop {} {\bibfield
  {journal} {\bibinfo  {journal} {Science}\ }\textbf {\bibinfo {volume}
  {372}},\ \bibinfo {pages} {1192} (\bibinfo {year} {2021})}\BibitemShut
  {NoStop}%
\bibitem [{\citenamefont {Rubio-Abadal}\ \emph {et~al.}(2020)\citenamefont
  {Rubio-Abadal}, \citenamefont {Ippoliti}, \citenamefont {Hollerith},
  \citenamefont {Wei}, \citenamefont {Rui}, \citenamefont {Sondhi},
  \citenamefont {Khemani}, \citenamefont {Gross},\ and\ \citenamefont
  {Bloch}}]{rubio2020floquet}%
  \BibitemOpen
  \bibfield  {author} {\bibinfo {author} {\bibfnamefont {A.}~\bibnamefont
  {Rubio-Abadal}}, \bibinfo {author} {\bibfnamefont {M.}~\bibnamefont
  {Ippoliti}}, \bibinfo {author} {\bibfnamefont {S.}~\bibnamefont {Hollerith}},
  \bibinfo {author} {\bibfnamefont {D.}~\bibnamefont {Wei}}, \bibinfo {author}
  {\bibfnamefont {J.}~\bibnamefont {Rui}}, \bibinfo {author} {\bibfnamefont
  {S.}~\bibnamefont {Sondhi}}, \bibinfo {author} {\bibfnamefont
  {V.}~\bibnamefont {Khemani}}, \bibinfo {author} {\bibfnamefont
  {C.}~\bibnamefont {Gross}}, \ and\ \bibinfo {author} {\bibfnamefont
  {I.}~\bibnamefont {Bloch}},\ }\href@noop {} {\bibfield  {journal} {\bibinfo
  {journal} {Physical Review X}\ }\textbf {\bibinfo {volume} {10}},\ \bibinfo
  {pages} {021044} (\bibinfo {year} {2020})}\BibitemShut {NoStop}%
\bibitem [{\citenamefont {Nandkishore}\ and\ \citenamefont
  {Gopalakrishnan}(2017)}]{nandkishore2017many}%
  \BibitemOpen
  \bibfield  {author} {\bibinfo {author} {\bibfnamefont {R.}~\bibnamefont
  {Nandkishore}}\ and\ \bibinfo {author} {\bibfnamefont {S.}~\bibnamefont
  {Gopalakrishnan}},\ }\href@noop {} {\bibfield  {journal} {\bibinfo  {journal}
  {Annalen der Physik}\ }\textbf {\bibinfo {volume} {529}},\ \bibinfo {pages}
  {1600181} (\bibinfo {year} {2017})}\BibitemShut {NoStop}%
\end{thebibliography}%

\appendix
\section{Rotating frame}
Quantum systems subjected to random multipolar driving (RMD) have been shown to exhibit long-lived prethermal quasisteady states, with lifetimes that grow algebraically with the driving rate $1/T$~\cite{mori2021rigorous,zhao2021random}.
This behavior is obtained when $1/T$ is large compared with the energy scales associated with all other local terms in the system's Hamiltonian.
However, 
the anomalous random multipolar driven insulator (ARMDI) introduced in the main text has one energy scale $J$ whose magnitude is locked to be proportional to $1/T$. 
Hence the previous results exhibiting long lifetimes for RMD systems do not directly apply here. 

To demonstrate the prethermal stability of the ARMDI, we define a rotating frame transformation the removes the dynamics associated with the hopping with energy scale $J \sim 1/T$:
\begin{eqnarray}
	Q(t)=\mathcal{T} e^{-i \int_{0}^{t} d s H_{\mathrm{id}}(s)},
\end{eqnarray}
where, as defined in Eq.~(1) of the main text,  $H_{\rm id}(s)$ generates the ``ideal'' hopping sequence in which each particle circles around one plaquette in time $T$.
Importantly, for systems with periodic boundary conditions (PBCs) one obtains $Q(T)=I$, as particles all hop back to their initial position after the first four steps. 

The new Hamiltonian in the rotating frame is given by
\begin{eqnarray}
	\tilde{H}^\pm(t)=Q^{\dagger}(t) {H^{\pm}}(t) Q(t)-i Q^{\dagger}(t) \dot{Q}(t),
\end{eqnarray} 
where $H^\pm(t) = H_{\mathrm{id}}(t)+H_{\mathrm{dis}}(t)+H_{\mathrm{pert}}^{\pm}(t)$ (see main text).
This transformation gives
\begin{eqnarray}
	\tilde{H}^{\pm}(t)=Q^{\dagger}(t)\left[H_{\mathrm{dis}}(t)+H_{\mathrm{pert}}^{\pm}(t)\right] Q(t).
\end{eqnarray}
Note that $H_{\rm id}$ has been canceled, and thus there are no terms of magnitude $J$ in $\tilde{H}^\pm(t)$. 
Also, as $Q(t)$ is equal to the identity during the fifth step of each block, $\tilde{H}_{\rm dis}(t)$ (which itself is only nonzero during the fifth step) remains unchanged in the new frame. 

Now we discuss the behavior of the perturbation in the rotating frame.
The specific form of the perturbation should not result in qualitatively different results, as long as it is local and its amplitude remains small. Here, as in the main text, we consider 
\begin{equation}
	\begin{aligned}
		&H_{\rm pert}^{\pm} (t) = \sum_{n=1}^{4} f_{n}(t)\left[  \delta J\sum_{\mathbf{r} \in A} \left(c_{\mathbf{r}+\mathbf{d}_{n}}^{\dagger} c_{\mathbf{r}}+\mathrm{h.c.}\right) {\pm} \delta h H_{\rm sub}\right], \\ &H_{\rm sub} =  \sum_{\mathbf{r}\in A}n_{\mathbf{r}}-\sum_{\mathbf{r}\in B}n_{\mathbf{r}},
	\end{aligned}
\end{equation}
where ${\bf d}_n$ is a nearest neighbor bond vector as defined in the main text.
The first term represents a deviation of the hopping amplitude from its ideal value (where each particle hops between neighboring sites with probability one during a given step). 
This term transforms to a new (next-nearest neighbor) hopping term in the rotating frame, remaining local on this scale.
The second contribution is a sublattice potential that is present during the hopping steps. 
Note that a nonzero $\delta J$ or $\delta h$ will both induce imperfect hopping during the first four steps. 
Here we set $\delta h\neq0$ and $\delta J=0$ for simplicity. In the rotating frame, we have new driving Hamiltonians for the $+$ and $-$ blocks given by
\begin{eqnarray}
	\label{eq.rotating_hamiltonian}
	\tilde{H}^{\pm} (t)= H_{\rm dis}f_5(t) \pm \delta h \sum_{n=1}^4 f_n(t) Q^{\dagger}(t)H_{\rm sub}Q(t).\ 
\end{eqnarray}
The second term, corresponding to the sublattice potential, involves the transformed operators $\tilde{n}_{\mathbf{r}}(t) \equiv Q^{\dagger}(t) n_{\mathbf{r}} Q(t)$.
Note that each operator $\tilde{n}_{\mathbf{r}}(t)$ has its support only on $\mathbf{r}$, as well as the nearest neighbor and next-nearest neighbor sites of $\mathbf{r}$~\cite{nathan2019anomalous}. To be precise, we take the result from  Ref.~\cite{nathan2019anomalous} for the first driving step as an example
\begin{eqnarray}
	\begin{aligned}
		\tilde n_{\mathbf{r}}(t)&=\cos ^{2}(J t) c_{\mathbf{r}}^{\dagger} c_{\mathbf{r}}+\sin ^{2}(J t) c_{\mathbf{r}+\sigma_{\mathbf{r}} \mathbf{d}_{1}}^{\dagger} c_{\mathbf{r}+\sigma_{\mathbf{r}} \mathbf{d}_{1}} \\&+\frac{i}{2} \sin (2 J t)\left(c_{\mathbf{r}}^{\dagger} c_{\mathbf{r}+\sigma_{\mathbf{r}} \mathbf{d}_{1}}-\text {h.c.}\right), \quad 0 \leq t< T / 5,
	\end{aligned}
\end{eqnarray}
where $\sigma_{\mathbf{r}}=1$ for ${\mathbf{r}}$ in sublattice $A$ and $\sigma_{\mathbf{r}}=-1$ for $\mathbf{r}$ in sublattice $B$. Terms with support on the neighboring sites of $\mathbf{r}$ appear, e.g., $c_{\mathbf{r}}^{\dagger} c_{\mathbf{r}+\sigma_{\mathbf{r}} \mathbf{d}_{1}}$ and $c_{\mathbf{r}+\sigma_{\mathbf{r}} \mathbf{d}_{1}}^{\dagger} c_{\mathbf{r}+\sigma_{\mathbf{r}} \mathbf{d}_{1}}$. At later times $T/5\leq t<T$, terms involving next-nearest-neighbor sites of ${\mathbf{r}}$ also appear, but $\tilde{H}^{\pm}(t)$ still remains local.  Clearly, the hopping amplitude $J$ now does not contribute to the norm of the rotated Hamiltonian $\tilde{H}^{\pm} (t)$. Expressions for the operator $\tilde{n}_{\mathbf{r}}(t)$ for $\delta J\neq 0$ can be similarly obtained. Therefore, in the rotating frame, a rapid driving regime required for the prethermalization is achieved for $h_{\mathrm{max}},\delta J,\delta h\ll T^{-1}$. 

One can furthermore estimate the heating rate induced by a random drive, in two different limits $T\to 0$ and $\delta h\to 0$. One can consider a simplified time-independent version of Eq.~\ref{eq.rotating_hamiltonian}: $H^{\pm}=H_{\mathrm{dis}}\pm \delta h H_{\mathrm{p}}$ where $H_{\mathrm{p}}$ appears as a random perturbation. For small $\delta h$  we have the following operator distance $||...||$ characterizing the deviation between the time evolution operators $U_{0}^{\pm}=\exp [-iT (H_{\mathrm{dis}} \pm \delta h H_{\mathrm{p}})]$ and $\exp (-iT H_{\mathrm{dis}} )$ ($H_{\mathrm{dis}}$ can be treated as an unperturbed effective Hamiltonian to approximate the transient dynamics at early times)~\cite{zhao2021random}:
$$\|U_0^{\pm} - \exp (-iT H_{\mathrm{dis}}) \| \sim O(\delta h T).$$
This distance can be used to estimate the deviation (or the heating rate) after a single evolution block. Thus we see that, for a single evolution block the two limits $T\to 0$ and $\delta h\to 0$ are indeed equivalent: both of them lead to vanishing deviations after a single block.

Now consider the evolution after $m$ evolution blocks, with $U_{0}^{\pm}$ selected randomly in each block. The distance between the true evolution operator for random driving (with $U_{0}^{\pm}$ selected randomly on each cycle) and the “ideal” localized evolution $\exp (-iT H_{\mathrm{dis}})$ accumulates linearly with the number of periods, $m$, becoming~\cite{zhao2021random}: 
\begin{eqnarray}
	\begin{aligned}
		\|\underbrace{U_{-} U_{+}U_{+} U_{-} \ldots}_{m \text { random blocks }}-\exp(-i H_{\mathrm{dis}} mT)\|\\ \sim O(mT \delta h)\sim O(t \delta h).
	\end{aligned}
\end{eqnarray}
While reducing $T$ gives an improvement of the deviation per evolution block (period), i.e., in stroboscopic time, at a fixed absolute time $t=mT$ the deviation does not scale with $T$. Hence taking $T\to 0$ has little effect on the long-time dynamics in absolute time. In contrast, taking $\delta h\to 0$ meaningfully leads to a prolongation of stability. Rigorous proofs and detailed discussions can be found in Ref.~\cite{zhao2021random,mori2021rigorous}; this argument can also be generalized to time-dependent Hamiltonians in the form of Eq.~\ref{eq.rotating_hamiltonian}. 

Now we turn to random multipolar driving, where the lifetime obtains a nontrivial dependence on $T$ and $\delta h$, both in stroboscopic time and in absolute time.  Specifically, the main advantage of using n-RMD is that, for any nonzero integer $n$, the heating rate is further suppressed by a power of the inverse driving period~\cite{mori2021rigorous}, and the operator distance for the time evolution at (absolute) time $t$ scales as $O(tT^{n})$. The dependence on $\delta h$ is complicated and model dependent, but most importantly the deviation generically scales with different powers of $\delta h$  and of $T$. The additional power-law suppression in $T$ is universal, implying a power-law increase of the lifetime in absolute time. This shows how genuinely long-lived prethermal states may be stabilized by random multipolar driving.

\section{Magnetization density}
Here we follow Ref.~\cite{nathan2017quantized} and define the magnetic density used to identify the topological property of the bulk.
Micromotion of particles in this system can be characterized via the orbital magnetization
\begin{equation}
	M(t)=\frac{1}{2}(\mathbf{r} \times \dot{\mathbf{r}}(t)) \cdot \hat{\mathbf{z}},
\end{equation} with $\dot{\mathbf{r}}(t)=-i[\mathbf{r}, H(t)]$. $M(t)$ is equivalent to the response of the Hamiltonian to an applied uniform magnetic field $B$, $M(t) = -\partial H(t) / \partial B$. The orbital magnetic density associated with each plaquette $p$ can be defined as
\begin{equation}
	m_{p}(t)=-\frac{\partial H(t)}{\partial \phi_{p}}, \quad \phi_{p}=\int_{p} d^{2} r B(\mathbf{r}),
\end{equation}
where $\phi_{p}$ represents the magnetic flux applied through the plaquette $p$. For a state $|\psi(t)\rangle$, one can define the time-averaged expectation value of an operator $\mathcal{O}(t)$ as
\begin{equation}
	\langle\mathcal{O}\rangle_{\tau} = \frac{1}{\tau} \int_{0}^{\tau} d t\langle\psi(t)|\mathcal{O}(t)| \psi(t)\rangle.
\end{equation}

According to Ampere's law on the lattice~\cite{nathan2017quantized}, if the particle density $\rho$ is stationary throughout the system over the time averaging interval, i.e., $\langle\dot{\rho}\rangle_{\tau}=0$, 
the time averaged current on the bond between neighboring plaquettes $p$ and $q$ equals the difference between the associated time-averaged  magnetic densities 
\begin{equation}
	\label{eq.ALaw}
	\left\langle I_{p q}\right\rangle_{\tau}=\left\langle m_{p}\right\rangle_{\tau}-\left\langle m_{q}\right\rangle_{\tau}.
\end{equation}
In our model, although time translation symmetry is broken, such stationary states can still be approximately achieved for $\tau=MT$ with integer $M$ during the long-lived prethermal regime before heating happens.
Particle density only significantly changes within a strip of width $D$ around the boundary of the filled region (as shown later in  Fig.~\ref{fig:current_supp}), where $D$ represents the localization length of the prethermal localized states. Therefore, at the distance $d$ from the boundary and for $d\gg D$, particle density changes exponentially small in the ratio $d/D$. Hence, all bond currents vanish in regions deep inside the droplet and the associated magnetic density becomes uniform. 

This uniform value of magnetic density is system size dependent for finite-size systems (finite size effect will be later discussed in Sec.~\ref{sec.finitesize}). For a plaquette $p$ at distance $d$ from the boundary, {in the case of the AFAI (which is stable in the long time limit)} one has $\lim_{\tau\to\infty}\left\langle m_{p}\right\rangle_{\tau} = \bar{m}_{\infty}+\mathcal{O}(e^{-d/D})$, with $\bar{m}_{\infty}$ denoting the value in the thermodynamic limit.
It has been shown in Ref.~\cite{nathan2017quantized} that $\bar{m}_{\infty}$ is quantized as the bulk topological order parameter for the anomalous Floquet Anderson insulators. For a non-vanishing $\bar{m}_{\infty}$, Ampere’s law (Eq.~\ref{eq.ALaw}) implies that the time-averaged orbital current $\lim_{\tau\to\infty}\left\langle I_C \right\rangle_{\tau}$ (see definition of $I_C$ in the main text) passing through a cut (the length of the cut needs to be larger than the localization length $D$) around the boundary of the droplet is also quantized and equals $\bar{m}_{\infty}$ up to a correction exponentially small in the localization length. See Ref.~\cite{nathan2017quantized} for more details.

As shown in the main text, for the aperiodic driving protocol, localization in our system has a finite prethermal lifetime $\tau_{\mathrm{pre}}$. Therefore, instead of employing $\bar{m}_{\infty}$ which needs infinitely long time average, one can consider a temporal average over a time window $\tau'$ such that $\tau'\ll \tau_{\mathrm{pre}}.$ In practice, we numerically compute the orbital current $\left\langle I_C \right\rangle_{T}$ averaged over each block of duration $T$ to identify the corresponding time-averaged magnetization density $\bar{m}_{T}(t) = \left\langle m_{p}\right\rangle_{T}(t)+\mathcal{O}(e^{-d/D})$ for plaquette $p$ deep in the droplet. Although $\bar{m}_{T}(t)$ now becomes time-dependent, as shown in Fig.~3 (b) in the main text, it remains approximately constant and quantized in the prethermal regime. 

\section{Density profiles for TM driving}

\begin{figure}
	\centering
	\includegraphics[width=0.45\linewidth]{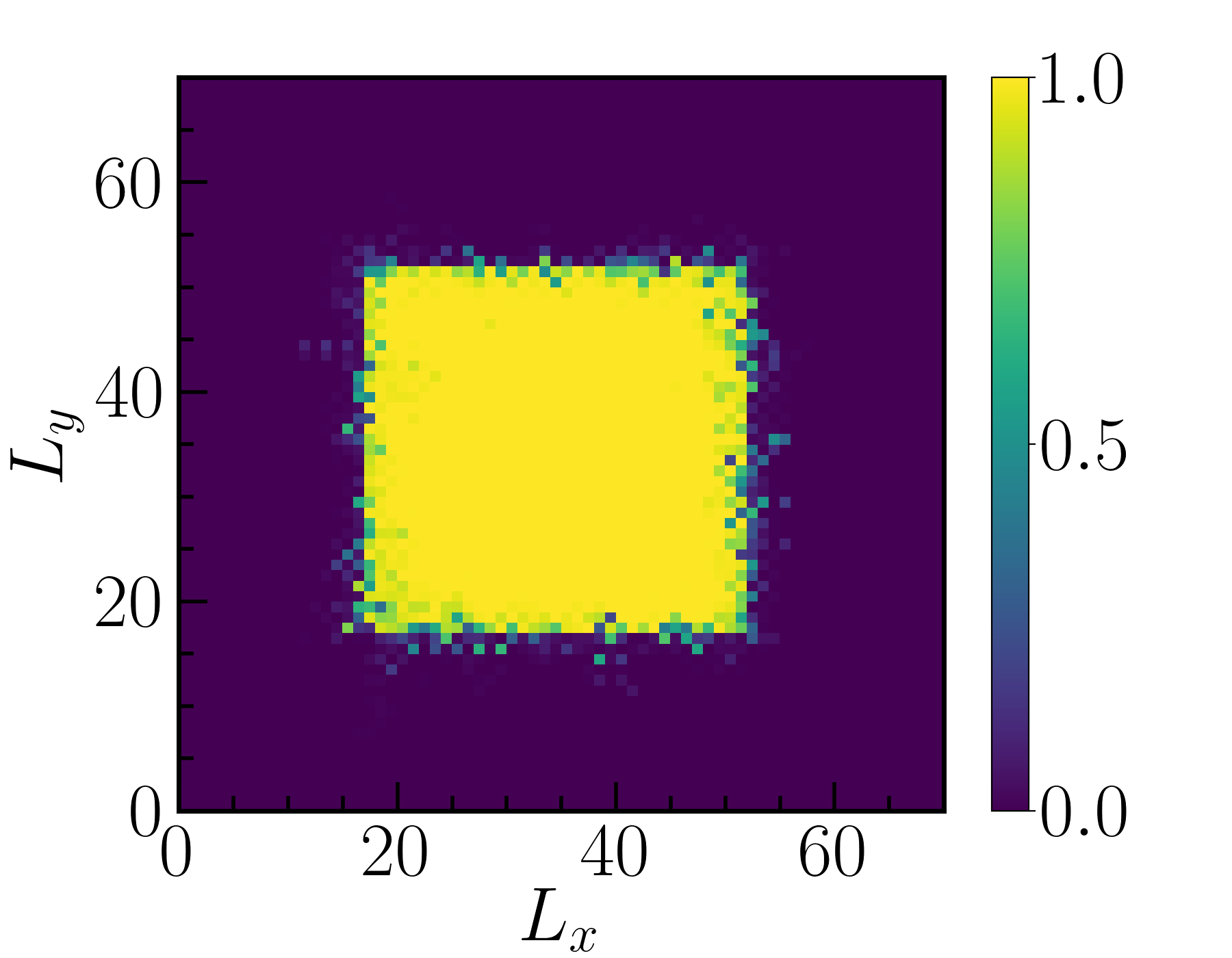}
	\includegraphics[width=0.45\linewidth]{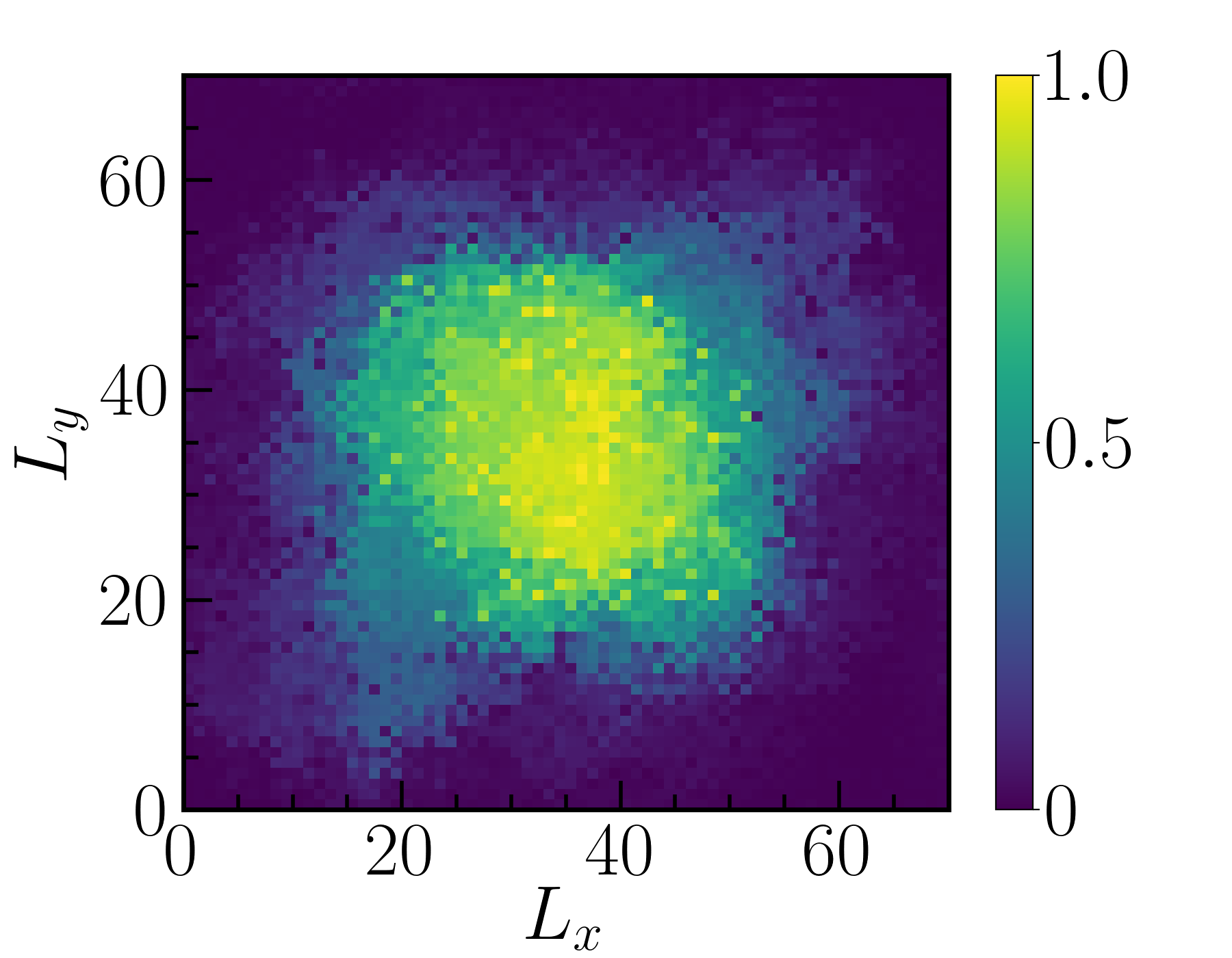}
	\includegraphics[width=0.45\linewidth]{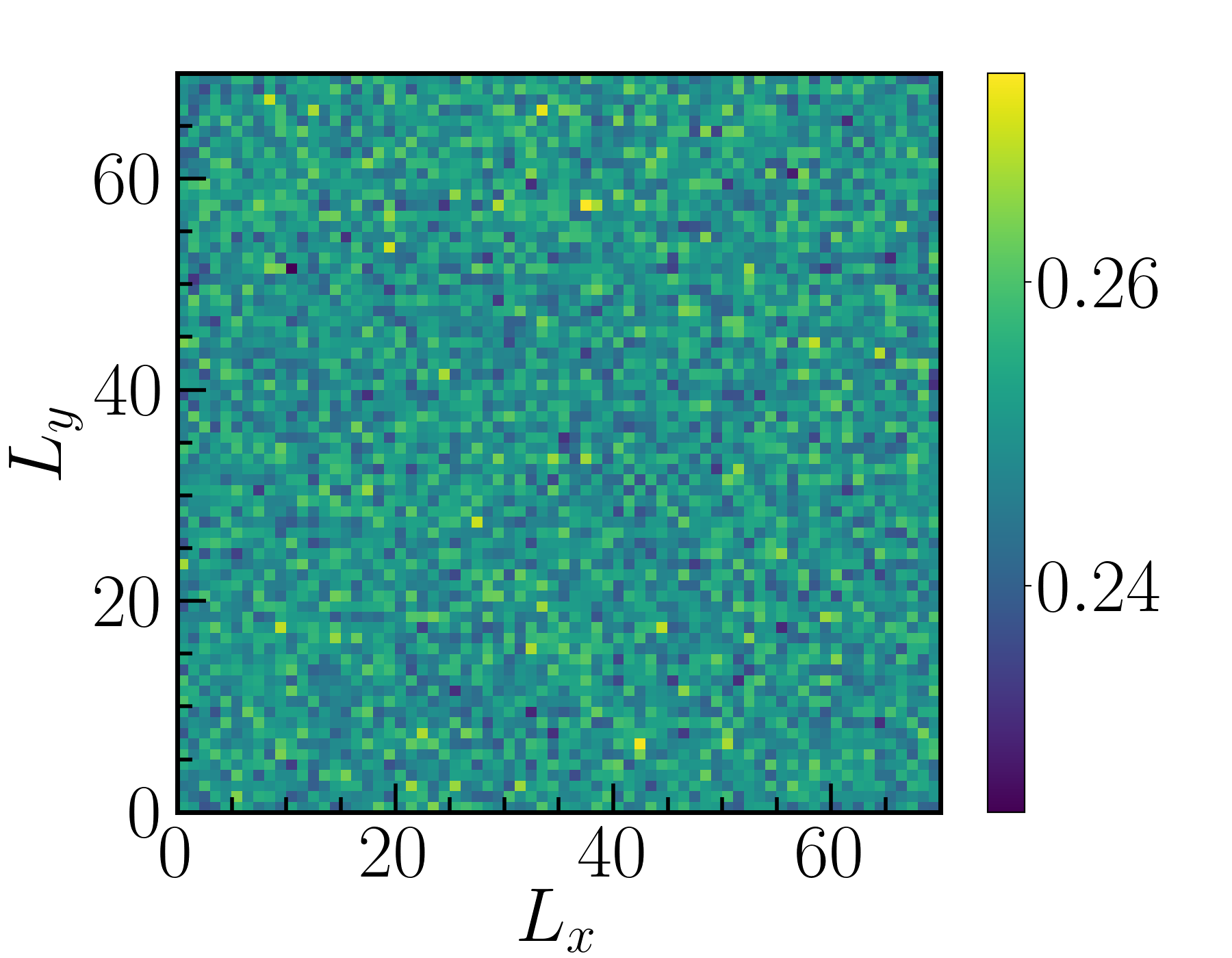}
	\caption{ Evolution of the particle density under Thue-Morse driving for a single random disorder realization for $1/T=20J_0$. The times shown in the three panels correspond to $2^{15}T, 2^{25}T, 2^{35}T$. 
		The parameters are $\delta J=1J_0,h_{\mathrm{max}}=20J_0,\delta h=7J_0.$}
	\label{fig:current_supp}
\end{figure}

Here we show additional density plots at different times using the same parameters as in Fig.~3 of the main text. As shown in Fig.~\ref{fig:current_supp}, the upper left panel shows the same results at $t\approx 10^3J_0^{-1}$ as in Fig. 3 (a) where the density remains almost the same as the initial distribution. The upper right panel shows the density around $t\approx1.5\times10^6J_0^{-1}$. 
Clearly the dynamics is not limited to the boundary of the square. A large region of the whole lattice has a nonzero density. After a sufficiently long time, e.g., $t\approx 2\times10^9J_0^{-1}$, the system exhibits a homogeneous distribution at particle density 0.25  indicating the eventual delocalization (bottom panel).

\section{Finite size effects}
\label{sec.finitesize}
Here we discuss the boundary effect which causes notable consequences to the lifetime of the prethermal localization. With periodic boundary conditions (PBCs), our model is similar to a localized system coupled to a thermal bath at infinite temperature~\cite{nandkishore2017many}.
The effect of the aperiodic drive on
Anderson localization can be treated as classical noise source. For open boundary conditions (OBCs), instead of being localized, states prepared at the boundary exhibit chiral propagation and quickly delocalize within a strip of the boundaries during the prethermal regime. Therefore, coupling to the delocalized boundary states further destablizes the Anderson localization in the bulk on top of the local random noise. This coupling  decays exponentially with distance. Hence such boundary effects can be well-controlled by either going to larger system sizes or stronger disorder strengths (corresponding to shorter bulk localization lengths). 

\begin{figure}
	\centering
	\includegraphics[width=\linewidth]{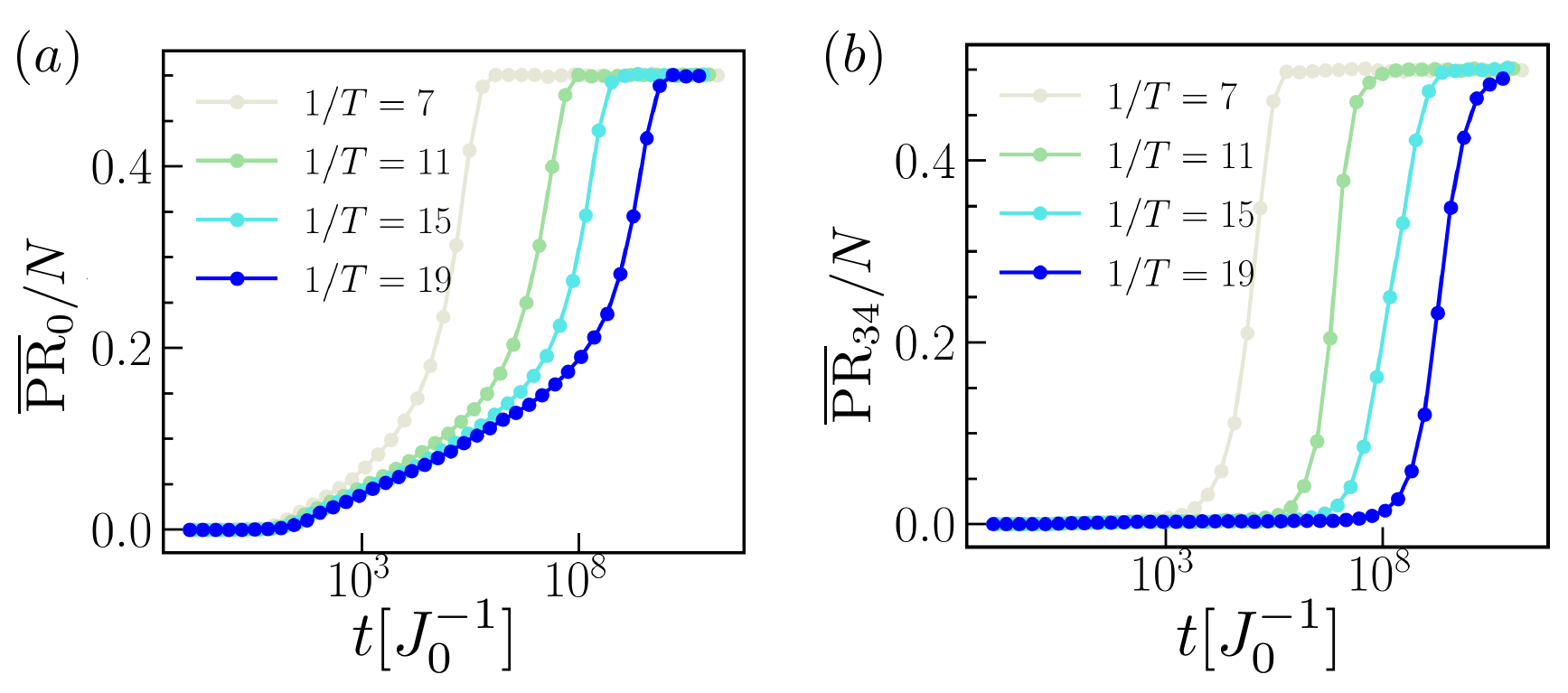}
	\caption{Dynamics of $\overline{\mathrm{PR}}_{L_i}/N$ for initial states prepared at $L_i$ sites away from the boundary of the cylinder of size $70\times70$. $L_i=0$ and 34 in panel (a) and (b), respectively. We use parameters $L_x=L_y=70,\delta J=1J_0,\delta h=7J_0, h_{max}=6J_0$. $T^{-1}$ is in the unit of $J_0$.}
	\label{fig:iprdynamicsobc1tms}
\end{figure}
\begin{figure}
	\centering
	\includegraphics[width=0.6\linewidth]{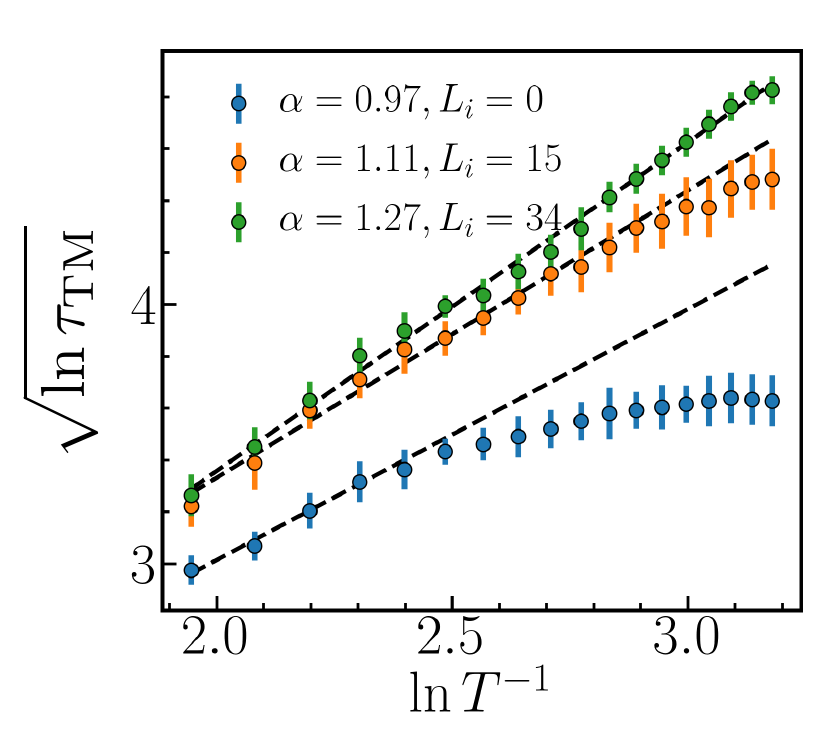}
	\caption{Scaling of the delocalization time versus driving rates for different $L_i$.  We use parameters $L_x=L_y=70,\delta J=1J_0,\delta h=7J_0, h_{max}=6J_0$. $T^{-1}$ is in the unit of $J_0$.}
	\label{fig:finitesizescaling}
\end{figure}
To demonstrate this phenomenon, in Fig.~\ref{fig:iprdynamicsobc1tms} we depict the evolution of the participation ratio $\overline{\mathrm{PR}}_{L_i}/N$ defined by
$$\overline{{\mathrm{PR}}}_{L_i} = \langle \mathrm{PR}_{m}\rangle_{L_i},$$
where the average is performed over all sites at distance $L_i$ from the top boundary of a cylinder of size $70\times 70$ and ${\rm PR}_m$ is as defined in the main text. Panels (a) and (b) of Fig.~\ref{fig:iprdynamicsobc1tms} show the result for $L_i=0$ and $L_i=34$, corresponding to the boundary and the center of the cylinder, respectively.  For $L_i=0$, the system first slowly delocalizes around $t\sim10^2J_0^{-1}$.
This onset time does not change for larger driving rate $1/T$. Dynamics of the participation ratio 
follow the form $\overline{\mathrm{PR}}_0/N\sim \log t$ over a large time window, e.g., from  $10^2J_0^{-1}$ to $10^8J_0^{-1}$ for $1/T=19J_0$, followed by a pronounced increase to the final plateau at 0.5. For $L_i$ deep in the bulk, as in panel (b), the boundary effect is negligible and a prethermal Anderson localization can be identified similarly to Fig.~2 of the main text, which was obtained with PBC.

The dependence of the delocalization time $\tau_{\mathrm{TM}}(L_i)$ on the driving rate $1/T$ and distance to the boundary, $L_i$, is plotted in Fig.~\ref{fig:finitesizescaling}. 
For each value of $L_i$, $\tau_{\mathrm{TM}}(L_i)$ is extracted by the procedure described in the main text as the average of times when $\overline{\mathrm{PR}}_{L_i}/N$ increases above the values $0.1\pm0.03,0.1\pm0.06$ for $L_i=34,15$, or $0.1\pm0.005,0.1\pm0.01$ for $L_i=0$.
The black dashed lines are described by the functional form  $\tau_{\mathrm{TM}} \sim e^{C[\ln (T^{-1}/g)]^{2}}$, and the fitted slope $\alpha$ corresponds to $\sqrt{C}$.  Note that the scaling exponent increases if the initial state is far from the edge, with a maximum value around 1.2. 
If we instead use a larger threshold value for determining $\tau_{\rm TM}(L_i)$, e.g., 0.4, a similar scaling exponent around 1 will be reproduced even for $L_i=0$. Such behavior
indicates that at the boundary, only the late stage of delocalization can be captured by the divergence of higher order operators of the TM sequence~\cite{mori2021rigorous}. 
As a comparison, for layers away from the open boundary, localization is stable and its lifetime scales similar as the system with PBC.

\begin{figure}
	\centering
	\includegraphics[width=\linewidth]{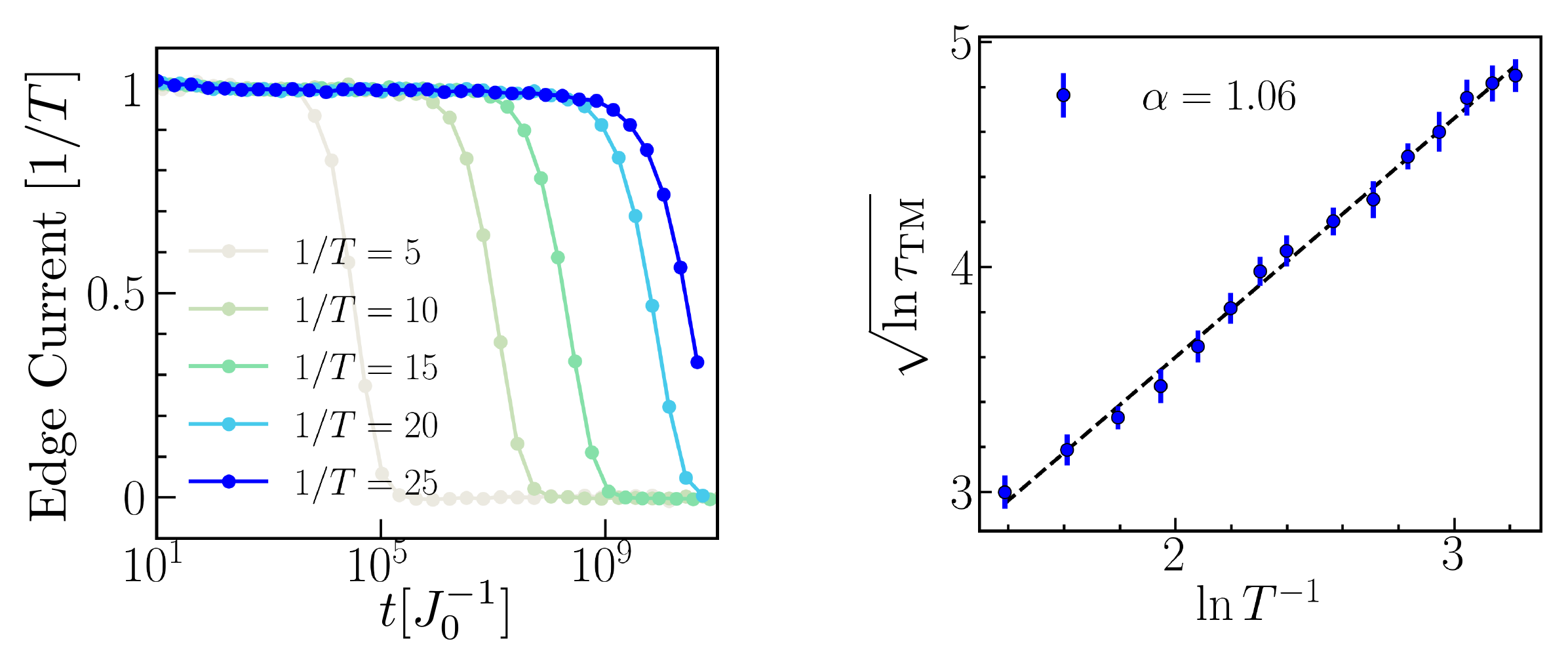}
	\caption{Dynamics of the edge current and its lifetime scaling for $L_x=L_y=70$. We use parameters $\delta J=1J_0,\delta h=7J_0, h_{max}=6J_0$ and $T^{-1}$ in the unit of $J_0$ .}
	\label{fig:Current_finitesize}
\end{figure}

Boundary effects also affect the scaling of the lifetime of edge current when we have a half-filled cylinder. The left panel in Fig.~\ref{fig:Current_finitesize} shows the dynamics of the (approximately) quantized current at the boundary between filled and empty sites.
The lifetime of this prethermal phenomenon is numerically extracted by averaging the times when the current drops below $0.7/T,0.7\pm0.2/T,0.7\pm0.1/T$. As shown in the right panel, the lifetime again fits well with $\tau_{\mathrm{TM}} \sim e^{C[\ln (T^{-1}/g)]^{2}}$, but the fitted slope ($\alpha$ corresponds to $\sqrt{C}$) is smaller than the largest value in Fig.~\ref{fig:iprdynamicsobc1tms} for $L_i=34$. This is reasonable as the deviation from the quantized current is induced by the delocalization around the central region of the lattice, for instance $L_i\in[L_y/2-\delta L, L_y/2]$, where $\delta L$ is a small finite integer and should be proportional to the localization length of the system.
As shown in Fig.~\ref{fig:iprdynamicsobc1tms}, for finite system sizes, the delocalization scaling exponent might still be dependent on $L_i$ and decrease for smaller $L_i$. Hence the scaling exponent for the current lifetime, which should involve contributions from layers of sites within $[L_y/2-\delta L, L_y/2]$, is slightly below the maximum delocalization scaling exponent.

\begin{figure}
	\centering
	\includegraphics[width=0.6\linewidth]{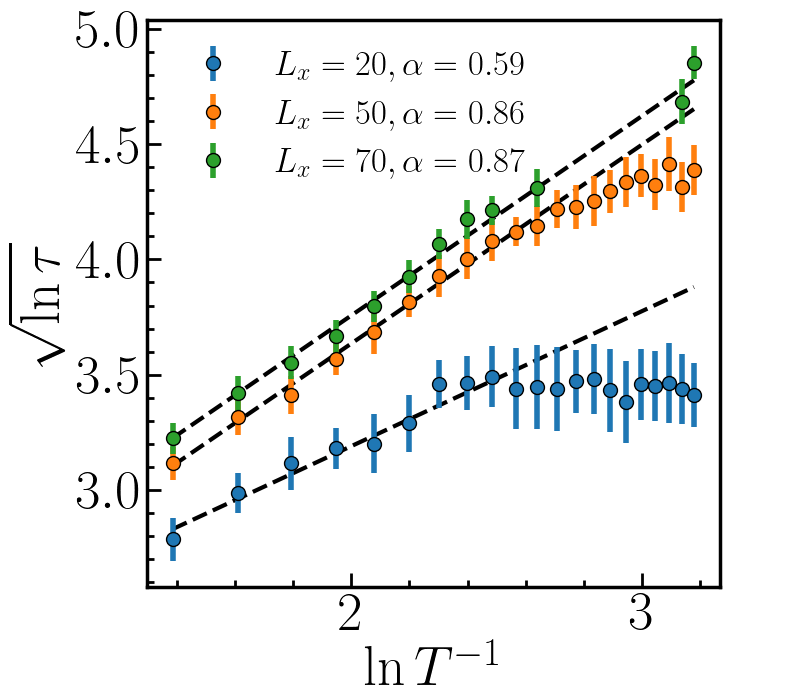}
	\caption{Scaling of the prethermal lifetime of edge current for different system sizes $L_x=20,50,70$. We use parameters $\delta J=1J_0,\delta h=5J_0, h_{max}=6J_0$ and $T^{-1}$ in the unit of $J_0$ .}
	\label{fig:Currentfinitesize}
\end{figure}

In Fig.~\ref{fig:Currentfinitesize} we also plot the prethermal life time of edge current for different system sizes. Deviations from the expected scaling $\tau_{\mathrm{TM}} \sim e^{C[\ln (T^{-1}/g)]^{2}}$ can be clearly observed for small system sizes at larger driving frequencies. But for sufficiently large system size, the scaling behavior converges for the numerically accessible time scales.

\section{RMD current}
\begin{figure}
	\centering
	\includegraphics[width=0.6\linewidth]{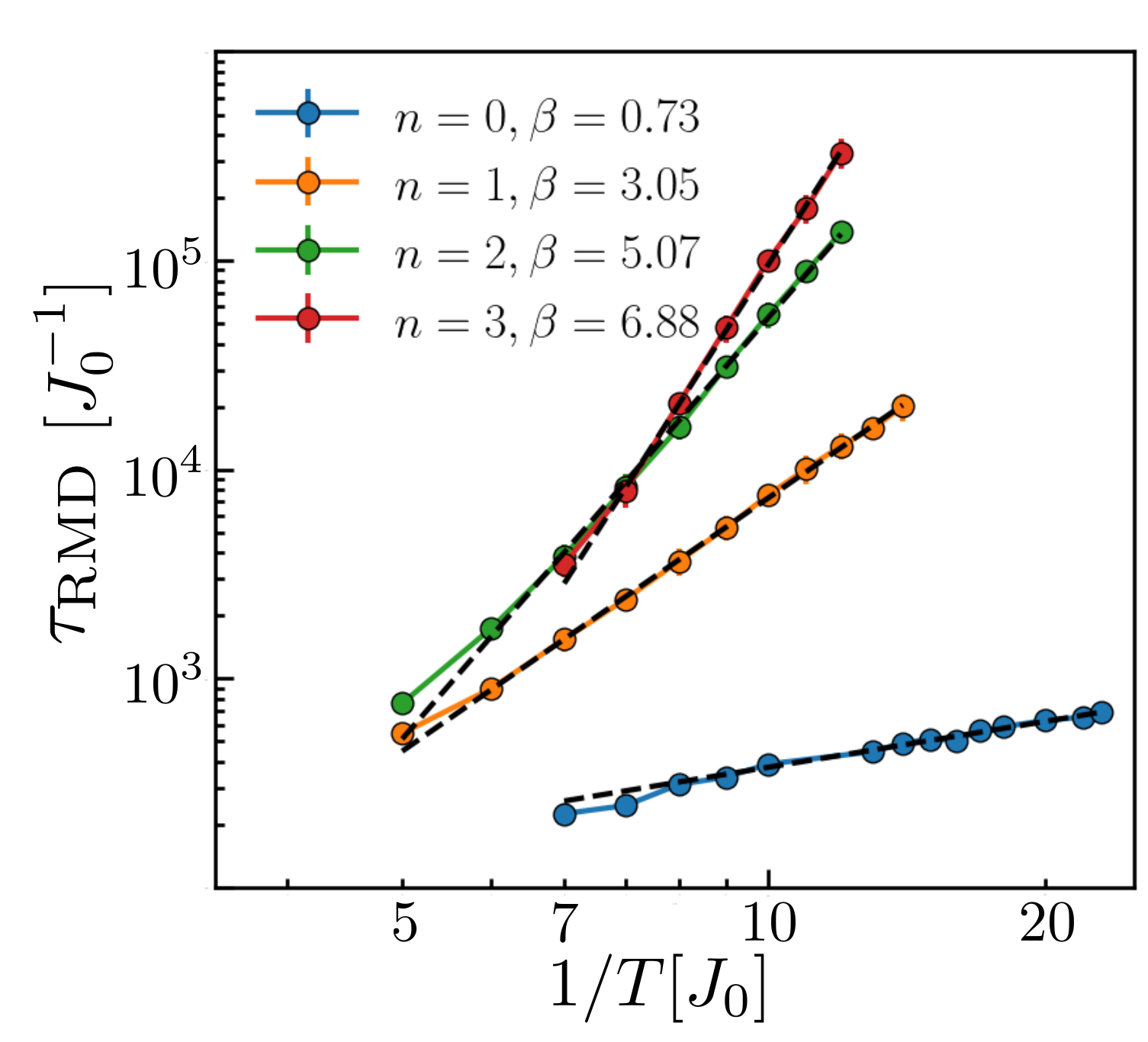}
	\caption{Dynamics of the edge current and its scaling for $n=0,1,2,3$ RMD. We use parameters $\delta J=1J_0, h_{max}=10J_0,\delta h=7J_0$ and system size $40\times 40$. }
	\label{fig:CurrentScaling_RMD}
\end{figure}
In Fig.~\ref{fig:CurrentScaling_RMD}, we show the scaling of the lifetime $\tau_{\mathrm{RMD}}$ for the ARMDI with different $n-$RMD protocols. The lifetime  is numerically extracted by averaging the times when the current drops below $0.96/T,0.96\pm0.015/T,0.96\pm0.0075/T$. The numerical results fit well with $\tau_{\mathrm{RMD}}\sim T^{-\beta}$ with $\beta\approx 2n+1$ for $n\ge 1$. Note that, compared with Fig.~2 of the main content, here we use a stronger disorder strength to reduce  finite size effects. Consequently, the lifetime obtained for purely random driving $n=0$ (blue dot in Fig.~\ref{fig:CurrentScaling_RMD}) also scales with the driving rate but with a very small exponent. We expect that for a weaker disorder and sufficiently large system size, the lifetime for $n=0$ should be independent of driving rate.

\end{document}